\documentclass[11pt]{article}
\usepackage{amsmath,amsfonts,amssymb,bbm,mathtools}
\usepackage{tensor}
\usepackage{slashed}
\usepackage[T1]{fontenc}
\usepackage[utf8]{inputenc}
\usepackage{lmodern}
\usepackage{authblk}
\usepackage{cite}
\usepackage{accents}
\usepackage{enumitem}
\usepackage[margin=3cm]{geometry}
\usepackage{varwidth}
\usepackage{hyperref}

\numberwithin{equation}{section}

\newcommand\nn{\ensuremath{\mathcal N}}
\newcommand\CC{\ensuremath{\mathbb C}}
\newcommand\RR{\ensuremath{\mathbb R}}
\newcommand\ZZ{\ensuremath{\mathbb Z}}

\newcommand\email[1]{\thanks{\href{mailto:#1}{\nolinkurl{#1}}}}

\DeclareMathOperator\tr{tr}
\DeclareMathOperator\U{U}
\DeclareMathOperator\OO{O}
\DeclareMathOperator\SO{SO}
\DeclareMathOperator\ISO{ISO}
\DeclareMathOperator\SU{SU}
\DeclareMathOperator\SL{SL}

\DeclareMathOperator\sign{sign}
\newcommand{\ket}[2]{\langle #1,#2\rangle}
\newcommand{\bra}[2]{[ #1,#2 ]}

\newcommand{\Pgen}[3]{\langle#1\!\mid\!#2\!\mid\!#3]}
\newcommand\st[1]{|#1\rangle}
\newcommand{\ubar}[1]{\underaccent{\bar}{#1}}
\newcommand{\Mhj}{M^{\phantom{j}}_{\textrm{ls}}\!\!\!\!\:\raisebox{4.7pt}{$\scriptstyle\{h_j\}$}}
\newcommand\Ac{\ensuremath{\mathcal{A}}}

\author[a,b]{Eduardo Conde\email{econdepe@snu.ac.kr}\,}
\author[b]{Andrea Marzolla\email{andrea.marzolla@ulb.ac.be}\,}
\newsavebox\affbox

\affil[a]{\protect\begin{varwidth}[t]{\linewidth}\protect\centering
School of Physics \& Astronomy and Center for Theoretical Physics,\par
Seoul National University, Seoul 08826, South Korea\protect\end{varwidth}\vspace{1.5ex}} 
\affil[b]{\protect\begin{varwidth}[t]{\linewidth}\protect\centering
Service de Physique Th\'eorique et Math\'ematique, Universit\'e Libre\par
de Bruxelles and International Solvay Institutes, Campus\par
de la Plaine, CP 231, B-1050 Bruxelles, Belgium\protect\end{varwidth}}

\title{\textsc{\LARGE 	Lorentz Constraints\\ \vskip 0.8ex
						on Massive Three-Point Amplitudes}}
\date{}
 
\begin{document}
\maketitle
\thispagestyle{empty}
\vskip 8em
\begin{abstract}
Using the helicity-spinor language we explore the non-perturbative constraints that Lorentz symmetry imposes on three-point amplitudes where the asymptotic states can be massive. As it is well known, in the case of only massless states the three-point amplitude is fixed up to a coupling constant by these constraints plus some physical requirements. We find that a similar statement can be made when some of the particles have mass. We derive the generic functional form of the three-point amplitude by virtue of Lorentz symmetry, which displays several functional structures accompanied by arbitrary constants. These constants can be related to the coupling constants of the theory, but in an unambiguous fashion only in the case of one massive particle. Constraints on these constants are obtained by imposing that in the UV limit the massive amplitude matches the massless one. In particular, there is a certain Lorentz frame, which corresponds to projecting all the massive momenta along the same null momentum, where the three-point massive amplitude is fully fixed, and has a universal form.
\end{abstract}

\cleardoublepage
\tableofcontents

\section{Invitation}

In four space-time dimensions the following algebraic accident occurs: the Lorentz group $\OO(3,1)$ is locally isomorphic to $\SL(2,\mathbb{C})$. The core of the helicity-spinor formalism resides in this simple observation, which allows to translate four-vectors into two-spinors, Lorentz indices into spinorial ones, and generically to recast $\OO(3,1)$ invariance as $\SL(2,\mathbb{C})$ invariance. Since the representation theory of the Lorentz group is typically constructed from the $\SL(2,\mathbb{C})$ one, then the spinorial language may sound like a good option for the study of the S-matrix, where representation theory plays a basal role.

Indeed, the adoption of the helicity-spinor formalism in part of the community of scattering amplitudes in the 80's led to more compact formulas, and to insights on a certain intrinsic simplicity of scattering amplitudes, commonly exemplified by the Parke-Taylor formula~\cite{Parke:1986gb}. By the end of the 90's this formalism established itself as the most convenient one for some practical purposes as the computation of QCD amplitudes (a good indicator may be the review~\cite{Dixon:1996wi}), and has also revealed extremely powerful for more theoretical purposes (many of them drawing inspiration from~\cite{Witten:2003aa}). Nowadays it permeates much of the work involving scattering amplitudes of four-dimensional QFTs, as can be grasped by looking at recent reviews, for instance~\cite{Dixon:2011xs,Elvang:2013cua}, or modern textbooks like~\cite{Srednicki:2007qs,Henn:2014yza}.

Here we want to emphasize another virtue of the helicity-spinor formalism, maybe not brought to light so often, which is that of simplifying the scattering-amplitude Ward identities associated to Lorentz symmetry. The latter follow from the transformation law of asymptotic one-particle states, which in turn is determined by the properties of infinite-dimensional representations of the Poincar\'e group. We define the basis of one-particle states through
\begin{equation}\label{Pst}
\hat{P}_{\mu}\st{k\,;a}=k_{\mu}\st{k\,;a} \ ,
\end{equation}
where $\hat{P}_{\mu}$ is the generator of translations and $k$ a four-momentum. The label $a$ stands for a collection of quantum numbers (for our purposes, mainly spin or helicity). Under a proper orthochronous Lorentz transformation $\Lambda\in\SO^+(3,1)$, these states generically transform as
\begin{equation}\label{Poincare}
U(\Lambda)\st{k\,;a}=\sum_{a'}D(\Lambda,k)_{aa'}\st{k';a'} \ ,\qquad k'=\Lambda k \ ,
\end{equation}
where the $D(\Lambda,k)_{aa'}$ furnish a representation of the Lorentz group. This transformation law induces a transformation law in the amplitude for the scattering of $n$ particles:
\begin{equation}\label{PoincareM}
M_n=\bigg(\prod_{i=1}^{n}\sum_{a'_i}D(\Lambda,k_i)_{a_ia'_i}\bigg)M'_n \ .
\end{equation}
In general the amplitude is not Lorentz invariant, it rather transforms covariantly. The form of the coefficients $D(\Lambda,k)_{aa}$ depends heavily on the character of $k$, \textit{i.e.} whether it is light-like or time-like (we will not consider the space-like case here). Because of that, we will use the following lower-case/upper-case notation throughout the paper:
\begin{equation}
k=p\quad\;\textrm{if}\;\; k^2=0 \ ,\qquad\qquad
k=P\quad\;\textrm{if}\;\; k^2<0 \ .\quad
\end{equation}
In the case of massless particles, the associated representation theory is quite simple. Barring internal quantum numbers, such as flavor, color, etc, the label $a=h$ is just an integer or half-integer called helicity. Moreover the coefficients $D(\Lambda,p)_{aa'}$ become diagonal, meaning that helicity is a Lorentz invariant. More precisely, $D(\Lambda,p)_{aa'}=e^{ih\theta(\Lambda,p)}\,\delta_{aa'}$, where $\theta(\Lambda,p)$ is the Little Group (LG) rotation angle induced by the Lorentz transformation $\Lambda$. Thus, equation~\eqref{PoincareM} recasts as
\begin{equation}\label{PoincareMm}
M_n=e^{ih_1\theta_1(\Lambda,p_1)}\cdots e^{ih_n\theta_n(\Lambda,p_n)}\,M'_n \ .
\end{equation}
The dependence of the $\theta_i(\Lambda,p_i)$ on $\Lambda$ and the external momenta is not particularly simple (for the explicit expression, see for instance~\cite{Weinberg:1995mt}, or~\cite{Schuster:2013pxj} in a nice more modern setup). So although this transformation law is quite compact, at first sight it is not evident at all what is the insight that we gain on the covariance properties of the amplitude. Or in other words, how much the transformation~\eqref{PoincareMm} constrains the functional form of $M_n$.

One possible way to make its meaning more transparent is to switch from LG indices (the helicities) to Lorentz ones. This is the approach Weinberg adopted in one of his celebrated works~\cite{Weinberg:1964ew}. Such a switch can be done via polarization tensors, which precisely carry out the task of transforming a LG index into a Lorentz one. The amplitude then becomes a Lorentz tensor, a ``tensor amplitude'', and it will transform as such. But there is a price to pay, which is gauge redundancy. The polarization tensors are not defined univocally, they do not transform covariantly, and extra requirements have to be imposed on the tensor amplitude (the so-called gauge Ward identities) so that equation~\eqref{PoincareMm} is verified.

However, if we decide to use the helicity-spinor formalism, equation~\eqref{PoincareMm} becomes more profitable, keeping LG indices instead of Lorentz ones. Using a pair of spinors to represent massless momenta as $p_i\!=\!\lambda_i\tilde\lambda_i$, the amplitude becomes simply a scalar function of these spinors $M_n\!=\!M_n(\lambda_{i},\tilde\lambda_i)$, in the sense that all spinor indices must be contracted. The amplitude satisfies the additional constraints
\begin{equation}\label{H_i}
\left(\lambda_{i}\frac{\partial^{\phantom{1}}}{\partial\lambda_{i}}-\tilde\lambda_{i}\frac{\partial^{\phantom{1}}}{\partial\tilde\lambda_{i}}\right)M_n=-2h_{i}\,M_n \ ,\qquad i=1,\ldots,n \ .
\end{equation}
These equations just guarantee that the amplitude carries the appropriate individual LG representations. In other words, individual copies of the Poincar\'e group act on the S-matrix as they act on one-particle states. Then the constraints above are just a consequence of the infinitesimal version of equation~\eqref{Poincare}.

The meaning of equations~\eqref{H_i} is very simple: the amplitude scales homogeneously with the spinors,
\begin{equation}
M_n(t_{i}\lambda_{i},t_{i}^{-1}\tilde\lambda_{i})=\prod_{i=1}^n t_{i}^{-2h_{i}}M_n(\lambda_{i},\tilde\lambda_{i}) \ .
\end{equation}
This condition surprisingly turns out to be so constraining that, for massless particles and $n\!=\!3$, it fixes all the functional dependence of the amplitude!\footnote{%
	Recall that scattering of three massless particles with real momenta is kinematically prohibited. However such an amplitude can be non-zero if we allow the momenta to be complex. Analyticity of the amplitude is assumed, and the analytic extension is done using equations~\eqref{H_i}.
}\cite{Benincasa:2007xk} The aim of this paper is to explore the effectiveness of analogous constraints on amplitudes involving massive particles as well. In other words, we want to see if Lorentz symmetry is powerful enough to fully determine the functional form of three-points amplitudes, independently of the details of the specific QFTs they come from. Such a question has been certainly considered in the past (see \textit{e.g.}~\cite{Strathdee:1967bi}, where this question was partially studied in an old-fashioned language), but, as far as we know\footnote{
More recently, several interesting works~\cite{Boels:2011zz,Boels:2012if} have also considered the question of Lorentz constraints on massive three-point amplitudes using the helicity-spinor formalism and extra ingredients, like supersymmetry or string theory. Our analysis does not require any of these ingredients and is therefore more general, although some of the amplitudes that we later discuss can be already found in those works.
}, not in full generality with the helicity-spinor language, which in our opinion greatly simplifies the analysis.

Although exploring some of the consequences of Poincar\'e symmetry for scattering amplitudes is interesting enough, one of our main motivations for this work comes from the success of the on-shell methods for massless particles based on recursion relations, like the BCFW recursion relations~\cite{Britto:2005aa}, where the three-point amplitude typically plays the role of seed. For instance, very powerful on-shell techniques, like the Grassmannian formulation of $\nn\!=\!4$~SYM~\cite{ArkaniHamed:2012nw}, contain as a small (but fundamental) ingredient the fact that the three-point massless amplitude is fully fixed by the symmetry of the theory. The recursion relations also work in the presence of massive particles~\cite{Badger:2005zh} (see~\cite{Badger:2005jv,Forde:2005ue,Ferrario:2006np,Boels:2010mj} for further references on massive BCFW recursion relations).  So if this seed can be determined from non-perturbative symmetry arguments, that can set up the stage for more profound discussions, noticeably in a scenario where several potential Infra-Red singularities are absent.

Actually, from related recent developments comes another of our motivations. When the Grassmannian techniques are employed for more realistic, less supersymmetric, theories, massive particles can pop up in the discussion, as a means to regulate forward limits~\cite{Benincasa:2015zna}. They appear as well in the scattering equations approach~\cite{Cachazo:2013hca}, where it has been recently conjectured that the key to extend the formalism to loop level lies in forward limits of scattering amplitudes involving massive particles~\cite{Geyer:2015bja,He:2015yua}.

\medskip

Let us now briefly sneak peek into the contents of this paper. We start in Section~\ref{helispin} by setting up the notation that will be used in the rest of the paper. We review the basics of the helicity-spinor formalism, and we apply it to massive particles by using the fact that a time-like momentum can always be decomposed into two light-like one. 

In Section~\ref{BC} we revisit the proof by Benincasa and Cachazo~\cite{Benincasa:2007xk} of the ``uniqueness'' of the massless three-point amplitudes, emphasizing certain particular aspects in preparation for the generalization to the massive case. Then in Section~\ref{massM3} we identify the appropriate spinor representations of the massive LG generators, drawing inspiration from~\cite{Chen:2011ve}, which allow to write and study the analogue of~\eqref{H_i} for massive particles.

In Section~\ref{solving} we show how to solve the equations for three-particle scattering amplitudes derived in Section~\ref{massM3}, focusing on the massive external states in their lowest spin projection. The solution does not come unique out of the box, but rather some undetermined functions remain. In Section~\ref{meet}, using some physical conditions, we show how these functions are actually just constants. We also study how massive amplitudes reduce to massless ones in the Ultra-Violet (UV) regime, which prompts us to analyze the reference frame where all the massive momenta are decomposed along the same null reference direction. In such a frame, the amplitude takes a fully fixed and universal form.

Before concluding with a summary and some remarks, we show in Section~\ref{pract} how our results can be matched with amplitudes computed via standard Lagrangian methods, for some emblematic processes involving massive particles. 

The reader who is interested in reproducing the computations will find in appendix~\ref{LG} some details concerning a technical study of the LG in the helicity-spinor formalism, and in appendix~\ref{raising} some formal derivations which underlie the results of Section~\ref{solving}.


\section{Setting up the notation: the helicity-spinor formalism}
\label{helispin}

Unfortunately helicity-spinor practitioners have not agreed on a single set of conventions. So in this section we are obliged to review well-known material just to make clear our notation, to be heavily used in the rest of the manuscript. Let us start by making precise the connection between the proper orthochronous Lorentz group in four dimensions, $\SO^+(3,1)$, and $\SL(2,\mathbb{C})$.

Using the Pauli matrices plus the identity, $\sigma^{\mu}\!=\!\left(\mathbbm{1},\sigma^1,\sigma^2,\sigma^3\right)$, we can associate a complex two-by-two matrix to any four-vector:
\begin{equation}\label{4to2x2}
\begin{array}{ccc}
\mathbb{R}^4 & \longrightarrow & M_2(\mathbb{C})\\
k_{\mu}=(k_0,k_1,k_2,k_3) & \longmapsto & \sigma^{\mu}_{a\dot{a}}k_{\mu}=\left(\begin{array}{cc} k_0+k_3 & k_1-ik_2\\k_1+ik_2 & k_0-k_3\end{array}\right)=k_{a\dot{a}}\;\;.
\end{array}
\end{equation}
This definition induces a map between $\SO^+(3,1)$, which acts on $\mathbb{R}^4$ as $k\mapsto \Lambda k$, and $\SL(2,\mathbb{C})$, which acts on $M_2(\mathbb{C})$ as $k\mapsto\zeta\,k\,\zeta^{\dagger}$. The map $\zeta(\Lambda)$ is given by:
\begin{equation}
\begin{array}{ccc}
\SO^+(3,1) & 
\longrightarrow & 
\SL(2,\mathbb{C})\\
\Lambda: k_{\mu}\mapsto \Lambda^{\nu}_{\mu} k_{\nu} & 
\longmapsto & 
\zeta(\Lambda):k_{a\dot{a}} \mapsto \zeta^b_a\,k_{b\dot{b}}\,{\zeta^{\dagger}}^{\dot{b}}_{\dot{a}}
\end{array}		\quad,\quad 
\begin{array}{c}
\zeta(\Lambda)^b_a k_{b\dot{b}}{\zeta^\dagger(\Lambda)}^{\dot{b}}_{\dot{a}} = \sigma^{\mu}_{a\dot{a}}\Lambda^{\nu}_{\mu}k_{\nu}\, \\
\Updownarrow \\
\zeta(\Lambda)^b_a \sigma^{\mu}_{b\dot{b}}{\zeta^{\dagger}(\Lambda)}^{\dot{b}}_{\dot{a}} = \sigma^{\nu}_{a\dot{a}}\Lambda^{\mu}_{\nu}
\end{array} \ .			\label{Ltransf}
\end{equation}
In order to have the Minkowski norm preserved, one should impose $k^2\!=\!(\Lambda k)^2$, obtaining $|\det(\zeta)|\!=\!1$. Given the way in which $\zeta$ acts, its phase has no physical relevance, so we are allowed to set $\det(\zeta)\!=\!1$. This is a left-over freedom, and any other choice would be equivalent. Note that, as Lie groups, the dimension of $\SL(2,\mathbb{C})$ is six, the same as $\SO^+(3,1)$, so the extra phase does not correspond to any physical transformation. The two groups are still not isomorphic though, as $-\zeta$ represents the same Lorentz rotation as $\zeta$. Quotienting by this transformation we do obtain an isomorphism: $\SO^+(3,1)\!=\!\SL(2,\mathbb{C})/\mathbb{Z}_2$. Indeed $\SL(2,\mathbb{C})$ is the universal cover of the Lorentz group.

Moreover we underline that the action of $\SL(2,\mathbb{C})$ on four-momenta, \emph{i.e.} $k\!\mapsto\!\zeta k \zeta^{\dagger}$, is crucial in order to map real momenta, \emph{i.e.} $(k_{a\dot{a}})^\dagger=k_{a\dot{a}}$, into real momenta. In the following, as it is usual and useful practice in the S-matrix program and recursion relations literature, we will deal with complex momenta. The complexified Lorentz group $\OO(3,1)_{\mathbb{C}}$ is locally isomorphic to $\SL(2,\mathbb{C})\times\SL(2,\mathbb{C})$, which acts on momenta as $k\mapsto\zeta\,k\,\tilde\zeta^{\dagger}$. Thus going to real momenta implies taking the diagonal in $\SL(2,\mathbb{C})\times\SL(2,\mathbb{C})$, that is: $\tilde\zeta\!\equiv\!\zeta$.

\smallskip

The definition~\eqref{4to2x2} determines how Lorentz contractions look like with spinorial indices
\begin{equation}
k\cdot l =\frac12\,\epsilon^{ba}\epsilon^{\dot{a}\dot{b}}\, k_{a\dot{a}} l_{b\dot{b}} \ .
\end{equation}
Naturally Lorentz invariants become $\SL(2,\mathbb{C})\times\SL(2,\mathbb{C})$ invariants now. It is then convenient to define two $\SL(2,\mathbb{C})$-invariant inner product for spinors, one for each representation of $\SL(2,\CC)$ under which they can transform:
\begin{equation}\label{innprod}
\ket{\kappa}{\lambda}\equiv\epsilon^{ba}\kappa_{a}\lambda_{b} \ ,\qquad\qquad
\bra{\smash{\tilde{\kappa}}}{\smash{\tilde{\lambda}}}\equiv\epsilon^{\dot{a}\dot{b}}
\tilde{\kappa}_{\dot{a}}\tilde{\lambda}_{\dot{b}} \ ,
\end{equation}
with $\epsilon^{12}=1=\epsilon^{\dot{1}\dot{2}}$, $\epsilon_{12}=-1=\epsilon_{\dot{1}\dot{2}}$, and $\epsilon^{ac}\epsilon_{cb}=\delta^{a}_{\phantom{a}b}$. These inner products are obviously anti-symmetric and, in particular, their vanishing implies that the spinors are proportional.

Another consequence of~\eqref{4to2x2} is that $k^2=-\det(k_{a\dot{a}})$. For a null momentum, $k=p$, the zero-determinant condition can be solved automatically by choosing
\begin{equation}\label{paa}
p_{a\dot{a}}=\lambda_a\tilde\lambda_{\dot{a}} \ .
\end{equation}
Instead, enforcing $\det(k_{a\dot{a}})=m^2$ for a massive momentum, $k=P$, cannot be done so cleanly. There are two standard ways of doing it, as discussed by Dittmaier~\cite{Dittmaier:1998nn}. We choose the strategy of representing $P$ as the sum of two null momenta (see~\cite{Boels:2009bv,Boels:2010mj,Cohen:2010mi,Boels:2011zz,Boels:2012if} for other interesting applications of this formalism):
\begin{equation}\label{Paa}
P_{a\dot{a}}=\lambda_a\tilde\lambda_{\dot{a}}+\mu_a\tilde\mu_{\dot{a}} \ ,\qquad
\ket{\lambda}{\mu}\bra{\tilde\lambda}{\tilde\mu}=-m^2 \ .
\end{equation}
Notice that the choices of spinors in~\eqref{paa} and~\eqref{Paa} are not unique. This non-uniqueness will be fundamental in the following sections, as the different choices can actually be connected via LG rotations, under which the amplitude has known transformation properties. Before closing this section, we specify two familiar shorthand notations that will be often used in the remainder of the paper. The first one is for the product of a pair of light-like momenta $p_{i}\!=\!\lambda_{i}\tilde\lambda_{i}$ and $p_{j}\!=\!\lambda_{j}\tilde\lambda_{j}$. In order to avoid writing $\lambda$'s all the time, we denote
\begin{equation}\label{pipj}
2\,p_{i} \cdot p_{j} = \langle\lambda_{i},\lambda_{j}\rangle[\tilde\lambda_{i},\tilde\lambda_{j}]\equiv \langle i,j\rangle\left[i,j\right] \ .
\end{equation}
The second shorthand notation is for the product of a light-like vector $p=\lambda\tilde\lambda$ with a generic four-momentum $k$:
\begin{equation}
2p\cdot k=\epsilon^{ba}\epsilon^{\dot{a}\dot{b}}\,\lambda_a\tilde\lambda_{\dot{a}}k_{b\dot{b}}\equiv-\Pgen{\lambda}{k}{\tilde\lambda} \ .
\label{Pgen}
\end{equation}
The minus sign on the last line has been introduced to make intuitive the reduction of equation~\eqref{Pgen} to~\eqref{pipj} in the particular case where $k\!=\!\kappa\tilde\kappa$ is light-like: 
\begin{equation}
-\Pgen{\lambda}{\kappa\tilde\kappa}{\tilde\lambda} = -\ket{\lambda}{\kappa}\bra{\tilde\kappa}{\tilde\lambda} = \ket{\lambda}{\kappa}\bra{\tilde\lambda}{\tilde\kappa} \ .
\end{equation}

\section{Revisiting the massless three-point amplitude}
\label{BC}

In a very nice piece of work~\cite{Benincasa:2007xk}, among many other things Benincasa and Cachazo realized that, with some physical requirements, equation~\eqref{H_i} fixes the three-point amplitude up to a coupling constant. In this section, we revisit both the passage from~\eqref{H_i} to~\eqref{PoincareM} and the solution of the former equation, in such a way that the transition from massless to massive particles later on be as smooth as possible.

\smallskip

In order to specify the form and transformation properties of one-particle states as in~\eqref{Poincare}, one has to find the infinite-dimensional unitary irreducible representations (irreps) of the Poincar\'e group acting on the Hilbert space of physical states. As Wigner showed long time ago, it is sufficient to study the representation theory of the subgroup that leaves the momentum of the particle invariant. This subgroup is called the Little Group, and depends on the Lorentz character of the momentum. Its irreps induce irreps of the Poincar\'e group.

For a massless particle with momentum $p$, the LG is $\ISO(2)$, the isometry group of $\mathbb{R}^2$. If we leave out of the discussion the so-called continuous-spin particles\footnote{Although they are interesting in their own right. See~\cite{Schuster:2013pxj,Schuster:2013vpr,Schuster:2013pta,Schuster:2014hca} for a recent reconsideration.}, we can consider that the LG is just $\U(1)$. This group is Abelian, and its irreps are very simple: they are characterized by an integer or half-integer number $h$, the helicity. Their defining property is
\begin{equation}\label{Jph}
J\,\st{p,h}=h\,\st{p,h} \ ,
\end{equation}
where $J$ is the infinitesimal generator of the LG. Of course $J$ is nothing but the generator of spatial rotations around the direction of movement of the particle, which indeed leaves $p$ invariant. In terms of the generators of spatial rotations, it can be written as $J\!=\!\frac{\vec{p}\cdot\vec{J}}{|\vec{p}|}$. Although $J$ has a trivial effect on $p\!=\!\lambda\tilde\lambda$, it acts non-trivially on the spinors. A Lorentz rotation $R=e^{i\theta\,J}$ rotates the spinors as
\begin{equation}
\lambda\to e^{-i\theta/2}\lambda \ ,\qquad\qquad\tilde\lambda\to e^{i\theta/2}\tilde\lambda \ .
\end{equation}
We gathered in the appendix \ref{LG.1} some details to gain physical intuition on this well-known formula. Here we just notice that indeed this transformation leaves $p\!=\!\lambda\tilde\lambda$ invariant, and that it preserves the reality of $p$, when $\tilde\lambda\!=\!\sign(p_0)\lambda^\dagger$. Actually, when we allow the momentum to be complex, the LG transformation is enlarged to $\lambda\to t\,\lambda$, $\tilde\lambda\to t^{-1}\tilde\lambda$, for generic $t\in\CC\setminus\!\{0\}$. It is now straightforward to write the generator $J$ in terms of spinor differentials:
\begin{equation}
J=-\frac{1}{2}\left(\lambda^a\frac{\partial}{\partial\lambda^a}-\tilde\lambda_{\dot{a}}\frac{\partial}{\partial\tilde\lambda_{\dot{a}}}\right)\equiv-\frac{1}{2}\left(\lambda\frac{\partial}{\partial\lambda}-\tilde\lambda\frac{\partial}{\partial\tilde\lambda}\right) \ .
\end{equation}
Plugging this expression in~\eqref{Jph} we get
\begin{equation}\label{Hp}
\left(\lambda\frac{\partial}{\partial\lambda}-\tilde\lambda\frac{\partial}{\partial\tilde\lambda}\right)\st{p;h}=-2h\,\st{p;h} \ .
\end{equation}
Tensoring this expression with the vacuum and other $n\!-\!1$ one-particle states, we obtain~\eqref{H_i}. Leaving a more detailed justification for Section \ref{masspol}, it is interesting to point out here that equation~\eqref{Hp} holds if we substitute $\st{p,h}$ by the polarization tensor of the state. Concrete realizations of the polarization tensors for a massless spin-$s$ particle can be found in~\cite{Benincasa:2007xk} for any value of $s$. They can be built out of tensoring appropriately those for spin-$\frac12$ and spin-$1$ particles, which read
\begin{equation}
\label{polsm0}
\begin{split}
\epsilon_{a}(p;-\frac12)&=\lambda_a \ ,\\
\epsilon_{\dot{a}}(p;+\frac12)&=\tilde\lambda_{\dot{a}} \ ,
\end{split}
\qquad\qquad
\begin{split}
\epsilon_{a\dot{a}}(p;-1)&=\frac{\lambda_a\tilde\mu_{\dot{a}}}{\sqrt{2}\bra{\tilde\lambda}{\tilde\mu}} \ ,\\
\epsilon_{a\dot{a}}(p;+1)&=\frac{\mu_a\tilde\lambda_{\dot{a}}}{\sqrt{2}\ket{\mu}{\lambda}} \ .
\end{split}
\end{equation}
The reference spinors $\mu$ and $\tilde\mu$ are completely arbitrary, and reflect the fact that polarization tensors are not univocally defined.

\smallskip

Let us now move to the solution of the equations~\eqref{H_i} for the massless three-particle amplitude. Denoting
\begin{equation}\label{Ji}
\qquad J^{\ubar{i}}=-\frac12\left(\lambda_{i}\frac{\partial}{\partial\lambda_{i}}-\tilde{\lambda}_{i}\frac{\partial}{\partial\tilde{\lambda}_{i}}\right) \qquad (\text{with no sum on } i),
\end{equation}
the three equations can be written as
\begin{equation}\label{JiM3}
J^{\ubar{i}}\,M^{\{h_{j}\}}\big(\lambda_{j},\tilde\lambda_{j}\big)=h_{i}\,M^{\{h_{j}\}}\big(\lambda_{j},\tilde\lambda_{j}\big) \ ,\quad i=1,2,3 \ .
\end{equation}
Notice that the amplitude depends not just on the momenta of the participating particles, but also on their states, \textit{i.e.} their helicities. We show it explicitly via a superindex, $M^{\{h_{j}\}}\!=\!M^{h_{1},h_{2},h_{3}}$, as it will be convenient for transitioning to the massive case. Because of Lorentz covariance, the three-point amplitude must be a function of the spinor products:
\begin{equation}
\left.\begin{array}{c}
x_1=\ket23 \ ,\;x_2=\ket31 \ ,\;x_3=\ket12\\
y_1=\bra23 \ ,\;\;y_2=\bra31 \ ,\;\;y_3=\bra12
\end{array}\right\}\; \Rightarrow \quad
M^{\{h_{j}\}}\big(\lambda_{j},\tilde\lambda_{j}\big)=M^{\{h_{j}\}}(x_j,y_j) \ .
\end{equation}
Using the chain rule, the operators~\eqref{Ji} are easily translated into
\begin{equation}
J^{\ubar{i}}=-\frac12\sum_{j\neq{i}}\left(x_j\frac{\partial}{\partial x_j}-y_j\frac{\partial}{\partial y_j}\right)\equiv-\frac12\sum_{j\neq{i}}\left(x_j\partial_j-y_j\tilde\partial_j\right) \ .
\end{equation}
The three equations~\eqref{JiM3} can be then recombined into
\begin{equation}
\label{eqsm0}
\begin{aligned}
\left(x_1\partial_1-y_1\tilde{\partial}_1\right) M^{\{h_{j}\}} 
& 	= \left(h_1-h_2-h_3\right) M^{\{h_{j}\}} \ , \\
\left(x_2\partial_2-y_2\tilde{\partial}_2\right) M^{\{h_{j}\}} 
& 	= \left(h_2-h_3-h_1\right) M^{\{h_{j}\}} \ , \\
\left(x_3\partial_3-y_3\tilde{\partial}_3\right) M^{\{h_{j}\}} 
& 	= \left(h_3-h_1-h_2\right) M^{\{h_{j}\}} \ .
\end{aligned} 
\end{equation}
Barring delta functions\footnote{One can argue that the cluster decomposition principle rules them out.}, the most general solution to these three equations is:
\begin{equation}\label{M3nom}
M^{\{h_{j}\}}=x_1^{h_1-h_2-h_3}x_2^{h_2-h_3-h_1}x_3^{h_3-h_1-h_2}f\left(x_1y_1,x_2y_2,x_3y_3\right) \ .
\end{equation}
It would have been completely equivalent to choose a pre-factor made of powers of the $y_i$. Our reason to display this formula somewhat asymmetrically will become apparent in Section~\ref{solving}.

Arriving here, we have assumed that all $x_i,y_i$ were independent. But this is not the case, because momentum conservation implies relations among them. We will discuss these relations in a general fashion in Section \ref{kine}. In the scenario with three massless particles they are quite simple: they impose that either $x_i\!=\!0$, or $y_i\!=\!0$. Clearly all the products $x_iy_i\!=\!0$, so there are only two classes of functions that yield a physically sensible amplitude, neither singular nor trivial:
\begin{equation}\label{2fs}
f=g_H \ , 	\quad\textrm{and}\quad 
f= g_A\: \left(x_1y_1\right)^{h_2+h_3-h_1} \left(x_2y_2\right)^{h_3+h_1-h_2} \left(x_3y_3\right)^{h_1+h_2-h_3} \ ,
\end{equation}
giving the following two possible three-point amplitudes\footnote{%
	In a parity-invariant theory the two amplitudes are related by conjugation, so that $g_{\textrm{H}}$ and $g_{\textrm{A}}$ are not independent.}%
\begin{align}
	\label{M3H}
	&M^{\{h_j\}}=g_{\textrm{H}}\,x_1^{h_1-h_2-h_3}x_2^{h_2-h_3-h_1}x_3^{h_3-h_1-h_2} \ ,\\
	\label{M3A}
	&M^{\{h_j\}}=g_{\textrm{A}}\,y_1^{h_2+h_3-h_1}y_2^{h_3+h_1-h_2}y_3^{h_1+h_2-h_3} \ .
\end{align}
To answer whether one is valid rather than the other, we resort to another physical condition. The three-particle amplitude must vanish for real kinematics, since a three-massless-particle scattering is kinematically prohibited. Since real kinematics implies $y_i\!=\!-x_i^*$, it must be that all $x_i,y_i\!=\!0$. Imposing $M^{\{h_j\}}(0,\ldots,0)\!=\!0$ forces us to choose~\eqref{M3H} for $h_1\!+\!h_2\!+\!h_3<\!0$ and~\eqref{M3A} for $h_1\!+\!h_2\!+\!h_3>\!0$. For the case $h_1\!+\!h_2\!+\!h_3=\!0$, the choice is ambiguous and actually the amplitude does not need to vanish, as (only) for this helicity configuration it is consistent to have the three real momenta aligned up along the same null direction~\cite{Adler:2016peh}. We are however not aware of any physical interaction that produces a three-particle process with $h_1\!+\!h_2\!+\!h_3=\!0$ (except of course for the trivial case of three scalars, where the amplitude is just a constant), so one may speculate that this case is somehow sick~\cite{Laure-Anne,McGady:2013sga}.

With this line of reasoning we have just reproduced the conclusions of~\cite{Benincasa:2007xk}, essentially that the three-particle amplitude is fixed up to a coupling constant. This conclusion is reached using no more than Poincar\'e symmetry plus sensible physical requirements, therefore it is a non-perturbative statement.

Nonetheless let us comment on one of the conditions we used, which was that of requiring the three-particle amplitude to be non-singular. That seems like a sensible requirement if we are talking about the full non-perturbative amplitude. However, at loop order in a perturbative expansion, singularities can show up. For instance taking a collinear limit in the $P_{(12)}\!=p_1\!+\!p_2$ channel of the one-loop three-vertex of YM with three negative-helicity gluons one gets the singular~\cite{Bern:1995ix}
\begin{equation}\label{all-}
M^{-1,-1,-1}_{\textrm{1-loop}}\propto\frac{\ket23\ket31}{\bra12}=\frac{\ket12\ket23\ket31}{\ket12\bra12} \ .
\end{equation}
Of course a singular object is not observable, and can only be handled with an appropriate regularization scheme. In a certain context\footnote{%
	Namely the equation~\eqref{all-} was needed in~\cite{Bern:2005hs} as a seed amplitude in a one-loop BCFW recursion relation for computing rational pieces of YM amplitudes.}%
$\,$it can make sense to call this object an ``amplitude'', but our point here is just to remark that it must obey the Lorentz transformation properties of an amplitude~\eqref{eqsm0}. Therefore the expression~\eqref{all-} must anyway satisfy the general form~\eqref{M3nom}. And indeed it does, if we choose $f\!=\!(x_3y_3)^{-1}$. What we want to illustrate with this example is that if we drop the requirement of finiteness of the amplitude, allowing it to be singular, then the function $f$ in~\eqref{M3nom} is not fixed \textit{a priori} anymore, and needs not correspond to any of the two options in~\eqref{2fs}.

We should conclude that in general~\eqref{M3nom} is a mathematically sound result. The further physical requirement that the amplitude be non-singular, as it should be at the non-perturbative level, or for a tree-level contribution, fixes the form of the otherwise arbitrary function $f$ to one of the two choices in~\eqref{2fs}.

\smallskip

\section{Little Group action on the massive three-point amplitude}
\label{massM3}

In this section we will mimic the logic that we used in the previous section, in order to study in spinorial form the analogue of equation~\eqref{H_i} for three-point amplitudes containing massive particles. That is, first we identify the LG of a massive particle. Then we find the defining property of its irreps in terms of the action of the LG generators, \textit{i.e.} the counterpart of~\eqref{Jph}. Finally, after expressing these generators in spinorial form, the homologue of equation~\eqref{H_i} will follow. The procedure is obviously valid for a generic $n$-point amplitude. We specify to $n=3$, which is the most easily constrained case.

Of course the LG of a massive particle with momentum
\begin{equation}\label{Pmass}
P=\lambda\tilde\lambda+\mu\tilde\mu \ ,\qquad
\ket{\lambda}{\mu}\bra{\tilde\lambda}{\tilde\mu}=-m^2 \ ,
\end{equation}
is $\SO(3)$. If we go to the rest frame of the particle, this group is just made of all the spatial rotations, which have no effect on $\vec{P}=0$. As $\SO(3)\!=\!\SU(2)/\mathbb{Z}_2$, and it is more common to talk about $\SU(2)$ representations, we commit a small abuse of language, and hereinafter we will talk about an $\SU(2)$ LG. The $\ZZ_2$ identification should be kept in mind, but it will not be very relevant for what follows. The three standard generators of $\SU(2)$ can be conveniently repacked as $J_+, J_0,J_-$, satisfying the algebra
\begin{equation}\label{algJa}
\big[J_+,J_-\big]=2J_0 \ ,\qquad\qquad \big[J_0,J_{\pm}\big]=\pm J_{\pm} \ .
\end{equation}
In the rest frame, they can be expressed in terms of the Lorentz generators as usual: $J_+\!=\!\hat{J}^{23}\!+\!i\hat{J}^{31}$, $J_0\!=\!\hat{J}^{12}$, $J_-\!=\!\hat{J}^{23}\!-\!i\hat{J}^{31}$. The irreps are characterized by a momentum and (this time) two integers or half-integers, $s$ and $\sigma$, with $\sigma\in\{-s,-s+1,...,s\}$. The generators act in the following way:
\begin{equation}\begin{aligned}
J_0\st{P;s,\sigma}&=\sigma\st{P;s,\sigma} \ ,\\
J_+\st{P;s,\sigma}&=\sqrt{(s-\sigma)(s+\sigma+1)}\st{P;s,\sigma+1} \ ,\\
J_-\st{P;s,\sigma}&=\sqrt{(s+\sigma)(s-\sigma+1)}\st{P;s,\sigma-1} \ .
\end{aligned}\label{Ja}\end{equation}
The number $s$ is the spin of the particle, and $\sigma$ is the projection of the spin on the $x^3$ direction. Only the former is Lorentz-invariant. The latter depends on the choice of Lorentz frame.

We find an unpleasant surprise when expressing the operators above in terms of spinor differentials (see appendix \ref{LG.2} for details). These generators correspond to the $\SU(2)$ acting as in~\eqref{Btransf}, which does not translate nicely in terms of spinor differentials. In order to escape this drawback, we will use a different, ``more covariant'', representation of the massive LG, which we will denote by $\widetilde{\SU}(2)$ in order to emphasize the distinction.

Consider the transformation
\begin{equation}\label{cina}
\left(\!\begin{array}{c}\lambda\\ \mu\end{array}\!\right)\to U\left(\!\begin{array}{c}\lambda\\ \mu\end{array}\!\right) \ ,\qquad
\left(\,\tilde\lambda \;\; \tilde\mu\,\right) \to \left(\,\tilde\lambda \;\; \tilde\mu\,\right)U^{\dagger} \ ,
\end{equation}
with $U\in\widetilde{\U}(2)$. This transformation leaves $P$ invariant
\begin{equation}
\lambda\tilde{\lambda}+\mu\tilde{\mu}=\Big(\,\lambda\;\;\mu\,\Big)\left(\!\begin{array}{c}\tilde{\lambda}\\\tilde{\mu}\end{array}\!\right)=\Big(\,\lambda\;\;\mu\,\Big)\,U^{\intercal} {U^{\intercal}}^{\dagger}\left(\!\begin{array}{c}\tilde{\lambda}\\\tilde{\mu}\end{array}\!\right) \ ,
\end{equation}
so the $\widetilde{\SU}(2)\subset\widetilde{\U}(2)$ is the perfect candidate to be identified with the LG. Indeed, let us denote the generators of this $\widetilde{\SU}(2)$ with a tilde as well: $\tilde{J}_+,\tilde{J}_0,\tilde{J}_-$ (they can be respectively identified with the Pauli matrices combinations $\frac{\sigma^1+i\,\sigma^2}{2}$, $\frac{\sigma^3}{2}$, $\frac{\sigma^1-i\,\sigma^2}{2}$). We show in the appendix~\ref{LG.2} how they are related to those of~\eqref{algJa} simply via a momentum-dependent Lorentz rotation. From~\eqref{cina} we can read their differential form, already written in~\cite{Chen:2011ve}.
\begin{equation}\label{tJs}
\begin{aligned}
&\tilde{J}_+=-\mu\frac{\partial}{\partial\lambda}+\tilde{\lambda}\frac{\partial}{\partial\tilde\mu} \ ,\\
&\tilde{J}_0=-\frac12\left(\lambda\frac{\partial}{\partial\lambda}-\tilde\lambda\frac{\partial}{\partial\tilde\lambda}-\mu\frac{\partial}{\partial\mu}+\tilde\mu\frac{\partial}{\partial\tilde\mu}\right) \ ,\\
&\tilde{J}_-=-\lambda\frac{\partial}{\partial\mu}+\tilde{\mu}\frac{\partial}{\partial\tilde\lambda} \ .
\end{aligned}
\end{equation}
This simple action on the spinors is what we meant above by a more covariant representation of the LG. The $\widetilde{\SU}(2)$ irreps are as in~\eqref{Ja}, just putting tildes everywhere. In Section~\ref{meet} we will come back to the meaning of the $\widetilde{\U}(1)\subset\widetilde{\U}(2)$ that has been left out of this discussion.

A natural question raises, about how $\st{P;s,\tilde{\sigma}}$ and $\st{P;s,\sigma}$ are related. Let us reason in the rest frame of the particle. Here $J_0\!=\!\hat{J}^{12}$ and $\sigma$ is just the projection of the spin on the $x^3$ direction. From the appendix \ref{LG.2}, we can see that if we choose $\lambda\tilde\lambda$ along the $x^3$ direction (and $\mu\tilde\mu$ spatially opposite to it), then\footnote{ %
	More precisely, with such a choice we see that $\alpha=a$ and $\beta=b$ in equation~\eqref{map} when $R=e^{i\theta J_0}$, so that $U=e^{i\tilde{J}_0}=R$}%
$\tilde{J}_0=J_0$, meaning that with this choice we can identify $\tilde{\sigma}\!\equiv\!\sigma$ and therefore the states $\st{P;s,\tilde{\sigma}}\!\equiv\!\st{P;s,\sigma}$. 
In general $\tilde{J}_0$ measures the spin projection along the spatial direction of $\lambda\tilde\lambda-\mu\tilde\mu$, and we will refer to $\tilde\sigma$ simply as spin projection.

\medskip

We are now ready to write an analogue of equation~\eqref{JiM3}, when massive particles are involved in a three-particle process. If particle $i$ is massless, with momentum $p_{i}=\lambda_{i}\tilde\lambda_{i}$, we can write for it
\begin{equation}
\label{tJi}
J^{\ubar{i}}\, \st{p_{i};h_{i}} = h_{i}\, \st{p_{i};h_{i}} \ ,
\end{equation}
with $J^{\ubar{i}}$ defined in~\eqref{Ji}. In the case particle $i$ is massive, $P_{i}=\lambda_{i}\tilde\lambda_{i}+\mu_{i}\tilde\mu_{i}$, we have instead the relations
\begin{equation}\label{tJa}
\tilde{J}^{\ubar{i}}_a\, \st{P_{i};s_{i},\tilde{\sigma}_{i}} = \jmath^{i}_a(s_{i},\tilde{\sigma}_{i})\, \st{P_{i};s_{i},\tilde{\sigma}_{i}+a} \ , \quad a=-1,0,1 \ ,
\end{equation}
where the $\tilde{J}_a$ are obtained dressing~\eqref{tJs} with the particle index $i$, and
\begin{equation}\label{jias}
\jmath^{i}_0(s_{i},\tilde{\sigma}_{i})=\tilde{\sigma}_{i} \ ,\qquad
\jmath^{i}_{\pm}(s_{i},\tilde{\sigma}_{i})=\sqrt{(s_{i}\mp\tilde{\sigma}_{i})(s_{i}\pm\tilde{\sigma}_{i}+1)} \ .
\end{equation}
The three-point amplitude will depend on all the labels characterizing the states, therefore on the spinors making up the momenta together with the helicity/spin labels. For compactness, let us omit the dependence on the former and display the dependence on the latter in the form of a superindex: $M^{\{\eta_{j}\}}=M^{\eta_1,\eta_2,\eta_3}$, where $\eta_{j}\!\equiv\!h_{j}$ in the case particle $j$ is massless, and $\eta_{j}\!\equiv\!\tilde{\sigma}_{j}$ in the case it is massive. With this notation, the equations for the amplitude look like
\begin{equation}\label{eqsM3}
\left\{	\begin{array}{ll}
J^{\ubar{i}}\,M^{\{\eta_{j}\}} = h_{i}\, M^{\{\eta_{j}\}} \phantom{\frac{}{\}}}
&	\textrm{if}\quad k_{i}^2=0	\\
\tilde{J}^{\ubar{i}}_a\,M^{\{\eta_{j}\}} =\jmath^{i}_a\, M^{\{\eta_{j}+\delta_{ij}a\}} \phantom{\frac{\big|}{}} 
&	\textrm{if}\quad k_{i}^2<0
\end{array}\right. \qquad i=1,2,3 \ .
\end{equation}
Opposite to what happened in the massless case~\eqref{JiM3}, these are not equations for a single function. Unless all massive particles are scalars, the equations rather mix amplitudes of different spin projection, \textit{i.e.} amplitudes $M^{\{\eta_{j}\}}$ with different values for $\eta_{i}\!=\!\tilde\sigma_{i}$, if the $i$-th particle is massive. Thus, a good strategy to study the system of equations~\eqref{eqsM3} is to focus on the amplitude with $\eta_{i}\!=\!-s_{i}$, that we will call, with a slight abuse of language, the ``lowest-spin'' amplitude. The generator $\tilde J^{\ubar{i}}_-$ kills this amplitude, and applying $\tilde J^{\ubar{i}}_+$ to it $2s_{i}$ times we should reach the ``highest-spin'' amplitude, which should be killed by $\tilde J^{\ubar{i}}_+$. Therefore we can consider the simplified set of equations
\begin{equation}
\label{eqsM3ls}
\left\{
\begin{array}{ll}
J^{\ubar{i}}\,M_{\textrm{ls}}^{\{h_{j}\}} = h_{i}\, M_{\textrm{ls}}^{\{h_{j}\}} \phantom{\frac{e}{\Big|}}
& 	\textrm{if}\quad k_{i}^2=0 	\\
\begin{array}{l}
\!\!\!		\tilde{J}^{\ubar{i}}_0\,M_{\textrm{ls}}^{\{h_{j}\}} = -s_{i}\, M_{\textrm{ls}}^{\{h_{j}\}}	\phantom{\Big|} \\
\!\!\!		\tilde J^{\ubar{i}}_-\,M_{\textrm{ls}}^{\{h_{j}\}}=0										\phantom{\Big|} \\
\!\!\!		\left(\tilde J^{\ubar{i}}_+\right)^{2s_{i}+1}\,M_{\textrm{ls}}^{\{h_{j}\}}=0
\end{array}
& 	\textrm{if}\quad k_{i}^2<0
\end{array}	\right. \qquad 
i=1,2,3 \ .
\end{equation}
This differential system involves a single function, the \emph{lowest-spin} three-point amplitude, that we denote as $\Mhj$ (notice that the lowest-spin condition only applies to the labels of the massive particles; the helicities $h_{j}$ of the massless particles, if any, are free). We will solve for this amplitude in Section \ref{solving}. The ``higher-spin'' amplitudes, where $\tilde\eta_{i}\!>\!-s_{i}$, can be obtained by simply applying $\tilde{J}^{\ubar{i}}_+$ to $\Mhj$, following the second equation in~\eqref{eqsM3}, with $a\!\equiv\!+1\,$.

As a side comment we remark here that if one is looking for a Lorentz-invariant ``amplitude'', the quantity that one should consider is the squared amplitude, averaged over spin components of the massive particles: $\sum_{\sigma_{i}}|M^{\{\eta_j\}}|^2$. It is only needed to sum over the spin projections of the massive particles involved. Using the fact that $\left(\tilde{J}_{a}\right)^*=-\tilde{J}_{-a}$ for the operators in~\eqref{tJs}, it is easy to prove that they all annihilate the averaged squared amplitude and thus this quantity is Lorentz invariant.

\subsection{Massive polarization tensors}			\label{masspol}

Before moving to the solution of the differential equations, let us make some remarks about polarization tensors, and in particular discuss how they look like for massive particles in spinor formalism.

The Poincar\'e group is made out of translations and Lorentz rotations. Although the combination of the two is not a direct product, when building Poincar\'e representations~\eqref{Pst} their roles are somewhat separated. The momentum label $k$ comes from the translations, while the internal labels $a$ come from the Lorentz rotations. Secretly, this separation is the same as the one we use when looking for free-particle solutions of a given wave equation\footnote{%
	Indeed one can build wave equations directly from the representation theory of the Poincar\'e group. Such a process is not unique though (neither are Lagrangians!), and such a road has not been much followed in modern times.
}. %
Generically such wave-functions look like
\begin{equation}
\psi(k,a)[x]=\epsilon(k\,;a)\,e^{-ik\cdot x} \ .
\end{equation}
The information about the properties of the wave-function regarding translations are encoded in the exponential factor. Instead, the Lorentz properties are carried by $\epsilon(k;a)$, for which we have conspicuously used a notation that makes it look like a polarization tensor. In the case of a spin-1 particle, this notation becomes the standard one. In the case of the spin-$\frac12$ electron of QED, the usual notation is $\epsilon(k\,;\pm1)\!=u_{\pm}(k),\,v_{\pm}(k)$. One could reasonably call $u$ and $v$ polarization spinors.

The point that we want to make here is that there is an obvious identification
\begin{equation}
\label{st=psi}
\langle x \st{k\,;a}\equiv\psi(k\,;a)[x] \ .
\end{equation}
In particular, the polarization tensors $\epsilon(k\,;a)$ should satisfy equation~\eqref{tJi} if $k$ is null or~\eqref{tJa} if $k$ is massive. For the next sections, it will be useful to have explicit expressions for polarization tensors of a massive particle with momentum~\eqref{Pmass} and spin $s=0,\frac12,...$ in terms of the spinors $\lambda,\tilde\lambda,\mu,\tilde\mu$.

For a scalar particle, evidently the polarization tensor is trivial. The first non-trivial case is spin $\frac12$. The only linear solutions to the equations~\eqref{tJa} are
\begin{equation}
\label{eDirac}
\begin{array}{c}
\epsilon(P;-\frac12)=\lambda\\
\epsilon(P;+\frac12)=-\mu
\end{array}
\qquad\textrm{or}\qquad
\begin{array}{c}
\epsilon(P;-\frac12)=\tilde\mu\\
\epsilon(P;+\frac12)=\tilde\lambda
\end{array} \ ,
\end{equation}
and which one to choose depends on the $\SL(2,\mathbb{C})$ representation we want the spinor to transform under. For a massive Dirac fermion, we have to combine the two representations:
\begin{equation}
\label{eDirac.bis}
\epsilon(P;-\frac12)=\left(\begin{array}{c} \tilde\mu \\ \pm\lambda \end{array}\right) \ ,\qquad
\epsilon(P;+\frac12)=\left(\begin{array}{c} \tilde\lambda \\ \mp\mu \end{array}\right) \ .
\end{equation}
By choosing the sign appropriately\footnote{The choice depends on conventions. One possibility, readily written with the massive helicity-spinor formalism we are adopting, can be found for example in appendix B of~\cite{Cohen:2010mi}.} we obtain the usual spinors $u$ and $v$, which solve the Dirac equation:
\begin{equation}
\left(\gamma^{\mu}P_{\mu}-m\right) v_{\pm}(P)=0 \ ,\qquad
\left(\gamma^{\mu}P_{\mu}+m\right) u_{\pm}(P)=0 \ .
\end{equation}
We can move now to the massive spin-1 particle, and follow the same steps. Asking for bilinear solutions of~\eqref{tJa}, up to numerical factors there is only one answer:
\begin{equation}
\label{pols1}
\epsilon(P;-1)=\lambda\,\tilde\mu \ ,\qquad \epsilon(P;0)=\frac{\lambda\tilde\lambda-\mu\tilde\mu}{\sqrt{2}} \ ,\qquad \epsilon(P;+1)=-\mu\,\tilde\lambda \ .
\end{equation}
If we do not require bilinear solutions, we have the freedom of incorporating arbitrary functions of $\ket{\lambda}{\mu}$ and $\bra{\tilde\lambda}{\tilde\mu}$ into the mix. We will do so in order to meet the requirement that the polarization vectors become those of~\eqref{polsm0} in the massless limit that we will discuss later in Section \ref{ssec:UV}. For the moment, we just take this limit as indicated in equation~\eqref{UVlimit}, and in order to recover the massless polarization vectors~\eqref{polsm0} we have to modify the expressions~\eqref{pols1} in the following way:
\begin{equation}
\label{pols1bis}
\epsilon(P;-1)=\frac{\lambda\,\tilde\mu}{\sqrt{2}\bra{\tilde\lambda}{\tilde\mu}} \ ,\qquad \epsilon(P;0)=\frac{\lambda\tilde\lambda-\mu\tilde\mu}{2m} \ ,\qquad \epsilon(P;+1)=-\frac{\mu\,\tilde\lambda}{\sqrt{2}\ket{\lambda}{\mu}} \ .
\end{equation}
These are the familiar\footnote{%
	It is immediate to check that, for real $P$, they verify the usual properties like $\epsilon(P;\pm1)^*\!\cdot\!\epsilon(P;\pm1)\!=\!1$, $\epsilon(P;\pm1)\!\cdot\!\epsilon(P;\pm1)\!=\!0$, and $\epsilon(P;\pm1)^*\!=\!\epsilon(P;\mp1)$.}
polarization vectors of a massive spin-$1$ particle. But if we want them to verify equations~\eqref{tJa}, we have to remember to consequently dress the $\tilde{J}_{\pm}$ with the factors
\begin{equation}
\label{rJ+}
\tilde{J}_+\to\sqrt{-\frac{\bra{\tilde\lambda}{\tilde\mu}}{\ket{\lambda}{\mu}}}\,\tilde{J}_+ \ ,\qquad
\tilde{J}_-\to\sqrt{-\frac{\ket{\lambda}{\mu}}{\bra{\tilde\lambda}{\tilde\mu}}}\,\tilde{J}_- \ .
\end{equation}
This rescaling leaves the algebra~\eqref{algJa} unchanged, and makes the $\tilde{J}_{\pm}$ ``$\mu$-helicity'' blind, a fact that we will take advantage of later on. Since for real momentum it is always possible to fix the factor inside the square root to 1, a natural way to think about formula~\eqref{rJ+} is as a physically sensible analytic continuation of $\tilde{J}_{\pm}$ to complex momenta.

As in the massless case, polarization tensors for spins $s\!>\!1$ can be constructed by tensoring the ones for $s\!=\!\frac12$ and $s\!=\!1$. Notice that simple factors of $\left(-\frac{\bra{\tilde\lambda}{\tilde\mu}}{\ket{\lambda}{\mu}}\right)^{\pm\frac12}$ must be included in the polarization spinors~\eqref{eDirac} to make the equations~\eqref{tJa} compatible with the rescaling~\eqref{rJ+}. For instance, for the Dirac fermion the polarization tensors that we will use later on are
\begin{equation}
\label{eDirac.bis.bis}
\epsilon(P;-\frac12)=\left(\begin{array}{c} \sqrt{-\frac{\ket{\lambda}{\mu}}{\bra{\tilde\lambda}{\tilde\mu}}}\,\tilde\mu \\ \pm\lambda \end{array}\right) \ ,\qquad
\epsilon(P;+\frac12)=\left(\begin{array}{c} \tilde\lambda \\ \mp\sqrt{-\frac{\bra{\tilde\lambda}{\tilde\mu}}{\ket{\lambda}{\mu}}}\,\mu \end{array}\right) \ .
\end{equation}

\smallskip

\section{Solving for the massive three-point amplitude}			\label{solving}

We now have almost all the tools to study the solutions of the differential equations~\eqref{eqsM3ls}. The only missing ingredient is the realization that, because of momentum conservation and on-shell conditions, the spinors that parametrize the momenta are not independent. Since the differential operators in~\eqref{eqsM3ls} commute with these constraints, we can take them into account before or after solving. In Section \ref{BC} we did it after, but due to the rapid growth in the number of variables as we include massive particles, now it will be more convenient for us to take these kinematic constraints into account from the outset. Therefore we start this section by first deriving the relations among the kinematic variables which the solution can depend on.

\subsection{Kinematic constraints}
\label{kine}

Since we decompose time-like momenta into two light-like ones, momentum conservation for the massive three-particle process looks like momentum conservation for a massless $n$-particle process, where $n$ will exceed $3$ by the number of involved massive particles. For convenience, for massive momenta $P_i=\lambda_i\tilde\lambda_i\!+\!\mu_i\tilde\mu_i$, we relabel $\{\mu_i,\tilde\mu_i\}\!\to\!\{\lambda_{j},\tilde\lambda_{j}\}$ with appropriate $j\!>\!3$ (we will specify the values of $j$ for each case in the subsections below). In this way, momentum conservation reads as:
\begin{equation}
\label{1+n}
\sum_{i=1}^n\lambda_i\tilde\lambda_i=0 \ .
\end{equation}
Out of the $2n$ spinors we can make $n(n-1)$ (angle  and square) products. We choose to keep most of the angle products as independent variables and to maximally reduce the number of independent square products. This simple systematic way to reduce the number of variables is motivated by our focus on the lowest-spin amplitude, but any other choice would be equivalent.

We first note that our spinors live in a 2-dimensional complex vector space, and so projecting onto two directions is enough to span the entire space:
\begin{equation}
\ket jk \lambda_i+\ket ki \lambda_j+\ket ij \lambda_k=0 \ .
\end{equation}
This fact goes under the name of Schouten identities. We can then choose for instance $\lambda_1$ and $\lambda_2$ as projecting directions, and express any of the angle products not involving neither $\lambda_1$ nor $\lambda_2$ in terms of
\begin{equation}
\ket{1}{2}\, , \;	\ket{1}{i}\, , \; 	\ket{2}{i}\, , \qquad \text{with } i=3,\ldots,\,n \ ,
\label{12i}
\end{equation}
which are $2\left(n\!-\!2\right)+1=2n\!-\!3$ independent variables. It is not possible to reduce the number of independent angle products using further Schouten identities: any Schouten involving $\ket12,\, \ket1i,\, \ket2j$ must also involve $\ket ij$.

So far we have only used constraints from linear algebra, and no momentum conservation. We will use the latter to reduce the number of square products. Contracting~\eqref{1+n} with $\lambda_1$ and $\lambda_2$ respectively, we obtain
\begin{equation}
\tilde\lambda_1=-\sum_{i=3}^n \frac{\ket{i}{2}}{\ket12}\tilde\lambda_i \ ,	\qquad \tilde\lambda_2=-\sum_{i=3}^n \frac{\ket{1}{i}}{\ket12}\tilde\lambda_i \ .	\label{momcocos}
\end{equation}
We have contracted with $\lambda_1$ and $\lambda_2$ to be consistent with the choice of~\eqref{12i}, and indeed we can see that only the angle products that we have chosen as independent variables in~\eqref{12i} appear in~\eqref{momcocos}. In addition, the Schouten identities among the angle products ensure that conditions~\eqref{momcocos} are sufficient to guarantee that momentum conservation be satisfied.

Thus we can derive in this way all the square products involving $\tilde\lambda_1$ or $\tilde\lambda_2$ in terms of
\begin{equation}
\bra{i}{j}	\quad 	\text{with }\, i,\,j\neq 1,2 \ ,		\label{braij}
\end{equation}
which are $\frac12(n-2)(n-3)$ variables. This number of variables can be further reduced to $2(n\!-\!4)+\!1=2n\!-\!7$ only if $n\!>\!5$, as we can still apply Schouten identities involving the left-over variables~\eqref{braij}. Schouten identities involving $\tilde\lambda_1$ or $\tilde\lambda_2$ follow straightforwardly from~\eqref{momcocos}, so they cannot contribute to further reduction of the number of independent square products.

So, according to our exploitation of momentum conservation and Schouten identities, we see that, out of the initial $n\left(n-1\right)$ spinor products, the number of independent ones is reduced to
\begin{equation}
\left\{
\begin{array}{ccccll}
2n-3 	& + &
\frac12(n-2)(n-3) 	& = &
\frac12 n\left(n-1\right) & 
\qquad \text{if }\, n\leq5 		\phantom{\Big|} \\
2n-3 	& + &
2n-7 	& = & 
2\left(2n-5\right) & 	
\qquad 	\text{if }\, n>5 		\phantom{\Big|}
\end{array}
\right. \ .			\label{nindvar}
\end{equation}

This conclusion is generic, and holds for any kinematic process with $n$ conserved massless momenta. In our setups, we have additional on-shell conditions, one per each massive particle. Applying these mass conditions we are able to finally reduce the counting to $6-1=5$ variables for one massive particle, $10-2=8$ variables for two massive particles, and $14-3=11$ variables for three massive particles. In the following we treat and solve separately these three cases.

\subsection{One massive leg}
\label{1m}

Let us consider the three-point amplitude involving one massive particle and two massless ones. The respective momenta will be $p_{1}$, $p_{2}$, and $P_{\ubar{3}}$, where we underline the index of the massive particle to avoid the mixing-up with the index of the spinors in the decomposition. Thus, in the spinor language we have
\begin{equation}
\label{p1234}
p_1=\lambda_1\tilde\lambda_1\ ,\qquad
p_2=\lambda_2\tilde\lambda_2\ ,\qquad
P_{\ubar{3}}=\lambda_3\tilde\lambda_3+\lambda_4\tilde\lambda_4 \ ,
\end{equation}
with the mass condition reading
\begin{equation}
\ket34 \bra43 =  {m_3}^2 \ .	\label{masscond}
\end{equation}
The system~\eqref{eqsM3ls} applied to the present case contains five equations. Let us start by solving the first four. Denoting the lowest-spin three-point amplitude as $M=M^{h_{1},h_{2},-s_{3}}$, they cast as
\begin{equation}
\begin{aligned}
\left(\lambda_1\frac{\partial}{\partial\lambda_1} - 
\tilde\lambda_1\frac{\partial}{\partial\tilde\lambda_1}\right) M 
=-2h_{1}\, M \ , \quad &
\left(\lambda_2\frac{\partial}{\partial\lambda_2} - 
\tilde\lambda_2\frac{\partial}{\partial\tilde\lambda_2}\right) M 
=-2h_{2}\, M  \ , \\
\left(\lambda_3\frac{\partial}{\partial\lambda_3} - 
\tilde\lambda_3\frac{\partial}{\partial\tilde\lambda_3} - 
\lambda_4\frac{\partial}{\partial\lambda_4} + 
\tilde\lambda_4\frac{\partial}{\partial\tilde\lambda_4}\right) M 
=+2s_{3}\, M \ , \quad &
\left(\lambda_3\frac{\partial}{\partial\lambda_4} - 
\tilde{\lambda}_4\frac{\partial}{\partial\tilde\lambda_3}\right) M 
=0  \ .
\end{aligned}					\label{eqs1Mleg}
\end{equation}
These equations can be easily translated into the language of spinor products, which we name for compactness as follows:
\begin{equation}
\label{12vars}
\begin{array}{clclclclclcl}
x_1=\ket23 	&\!\!\!,& x_2=\ket31 &\!\!\!,& x_3=\ket12 &\!\!\!,& x_4=\ket34 &\!\!\!,& x_5=\ket24 &\!\!\!,& x_6=\ket14 &\!\!\!, \\
y_1=\bra23 	&\!\!\!,& y_2=\bra31 &\!\!\!,& y_3=\bra12 &\!\!\!,& y_4=\bra34 &\!\!\!,& y_5=\bra24 &\!\!\!,& y_6=\bra14 &\!\!\!.
\end{array}
\end{equation}
We just need to use the chain rule to rewrite equations~\eqref{eqs1Mleg} in terms of these new variables (using the shorthand notation $\partial_{y_j}\!\equiv\!\tilde{\partial}_j$) as
\begin{equation}\label{ODE1m}
\begin{aligned}
\left(x_2\partial_2+x_3\partial_3+x_6\partial_6-y_2\tilde\partial_2-y_3\tilde\partial_3-y_6\tilde\partial_6\right) M 
&=-2h_{1}\, M \ , \\
\left(x_1\partial_1+x_3\partial_3+x_5\partial_5-y_1\tilde\partial_1-y_3\tilde\partial_3-y_5\tilde\partial_5\right) M 
&=-2h_{2}\, M  \ ,	\\
\left(x_1\partial_1+x_2\partial_2-x_5\partial_5-x_6\partial_6-y_1\tilde\partial_1-y_2\tilde\partial_2+y_5\tilde\partial_5+y_6\tilde\partial_6\right) M 
&=2s_{3}\, M	 \ , \\
\left(x_2\partial_6-x_1\partial_5+y_5\tilde\partial_1-y_6\tilde\partial_2\right) M &=0   \ .
\end{aligned}
\end{equation}
Now, we know that the variables~\eqref{12vars} are not all independent; so we apply the constraints as explained in the previous section, selecting as independent variables $x_1,\,x_2,\,x_3,\,x_4,\,x_5$. Given that the relations among variables are easy in this case, we display them explicitly for completeness:
\begin{equation}
\label{1m.cons}
\begin{array}{lll}
\displaystyle 	y_3=-\frac{m^2}{x_3} \ ,\, &
\displaystyle 		y_4=-\frac{m^2}{x_4} \ ,\, &
\displaystyle 			y_1=-\frac{m^2x_6}{x_3x_4} \ ,\, \\[2ex]
\displaystyle 	y_5=-\frac{m^2x_2}{x_3x_4} \ ,\, &
\displaystyle 		y_2=-\frac{m^2x_5}{x_3x_4} \ ,\, &
\displaystyle 			y_6=-\frac{m^2x_1}{x_3x_4} \ ,\,
\end{array}\qquad
x_6=-\frac{x_3x_4+x_2x_5}{x_1} \ .
\end{equation}
Using these constraints the system can be consistently\footnote{The derivatives appearing in~\eqref{ODE1m.2} should be understood as total derivatives, and the fact that the operators in~\eqref{ODE1m} can be expressed in terms of total derivatives follows from the commutation of the differential operators and our constraints. For example, the problem of solving $(x\partial_x+y\partial_y)f=3f$ subject to the constraint $y=x^2$ is ill-defined because the operator $(x\partial_x+y\partial_y)$ cannot be written in terms of $D_x=\partial_x+2x\partial_y$.} simplified to
\begin{equation}
\label{ODE1m.2}
\begin{aligned}
\left(x_2\partial_2+x_3\partial_3\right)M 	& = -2h_{1} M\ , \\
\left(x_1\partial_1+x_3\partial_3+x_5\partial_5\right)M		&=-2h_{2} M\ , \\ \left(x_1\partial_1+x_2\partial_2-x_5\partial_5\right)M 	& = +2s_{3}M\ , \\
\partial_5 M &=0\ .
\end{aligned}
\end{equation}
The very last equation tells us that the amplitude does not depend on $x_5$, so that the other three equations yield exactly the same system as in the massless case (with $h_{3}$ replaced by $-s_{3}$)! Moreover, we note that $x_4$ is not appearing in the equations, so that there is no constraint at all on the dependence of the amplitude on that variable. Thus, the most general solution for the one-massive-leg lowest-spin three-point amplitude is
\begin{equation}
\begin{aligned}
M^{h_{1},\,h_{2},-s_{3}} = \; &
x_1^{h_{1}-h_{2}+s_{3}}\, x_2^{h_{2}-h_{1}+s_{3}}\, x_3^{-s_{3}-h_{1}-h_{2}} \: f_1(x_4) \\
= \; &	{\ket12}^{-s_{3}-h_{1}-h_{2}}\: {\ket23}^{h_{1}-h_{2}+s_{3}}\: {\ket31}^{h_{2}-h_{1}+s_{3}}\; f_1\big(\ket34\big) \, ,
\end{aligned}	 	\label{sol1m}
\end{equation}
where $f_1$ is an arbitrary function, which depends on $\ket34$ and on other parameters of the interaction like the mass $m_{3}$, the coupling constant, etc. We can notice that the mass dimension of $f_1$ is fixed since the three-particle amplitude must have mass dimension equal to 1. Since $\ket34$ is dimensionful, it is convenient to write $f_1$ as
\begin{equation}
\label{tf1}
f_1(\ket34)=g\,m_{3}^{1+h_{1}+h_{2}-s_{3}-[g]}\tilde{f}_1\left(\frac{\ket34}{m_{3}}\right) \ ,
\end{equation}
where $g$ is the coupling constant controlling the interaction and $[g]$ is its mass dimension. Now the mass dimensions are provided by the $m_{3}$-dependent pre-factor, and the function $\tilde{f}_1$ is dimensionless, depending only on a dimensionless argument. In Section~\ref{meet} we will argue that the function $\tilde{f}_1$ should be constant.

We have not yet used the last equation in the system~\eqref{eqsM3ls}. The action of the raising operator $\tilde{J}_+^{\ubar{3}}$ on the amplitude~\eqref{sol1m} is spelled out in the appendix~\ref{raising}. Concretely, from equation~\eqref{J3nM1} we see that 
\begin{equation}
\label{J++1m}
\left(\tilde{J}^{\ubar{3}}_+\right)^{2s_{3}+1}M^{h_{1},\,h_{2},-s_{3}}=0\quad\implies\quad
(h_{1}-h_{2}-s_{3})_{(2s_{3}+1)}=0 \ ,
\end{equation}
where we are using the Pochhammer symbol
\begin{equation}
\label{Pochhammer}
(a)_{(n)}=a(a+1)\cdots(a+n-1) \ .
\end{equation}
Condition~\eqref{J++1m} means that the difference of the helicities can only take certain values, namely
\begin{equation}
\label{h1-h2}
h_{1}-h_{2}=-s_{3},-s_{3}+1,\ldots,s_{3} \quad \implies\quad
|h_{1}-h_{2}|\leq s_{3}\ .
\end{equation}
This is a strong constraint on the possible interactions with one massive particle, and it has a clear physical meaning: it comes out of the conservation of angular momentum. Since this process is allowed for real kinematics, we can place ourselves in a frame where the spatial momenta of the three particles are aligned. Imposing the conservation of momentum and angular momentum in this frame where the projections of the spin are maximized\footnote{
	In such a frame we can write $\frac{\vec{P}_{\ubar{3}}}{|\vec{P}_{\ubar{3}}|}=\pm\frac{\vec{p}_1}{|\vec{p}_1|}=\mp\frac{\vec{p}_2}{|\vec{p}_2|}$. We then have to multiply the first equality by $\vec{J}_1=\vec{J}_{\ubar{3}}-\vec{J}_2$, and use the second equality, to have the job done.
}, we see that the inequality~\eqref{h1-h2} is saturated. It is quite amusing to see how this physical condition comes out of the equations we wrote in~\eqref{eqsM3ls}.

\subsection{Two massive legs}
\label{2m}

We now repeat the procedure to find the solution for the three-point amplitude involving two massive particles and one massless one. We take particles 1 and 2 to be massive. The involved momenta will be $P_{\ubar{1}}$, $P_{\ubar{2}}$, and $p_3$, with
\begin{equation}
\label{p12345}
P_{\ubar{1}}=\lambda_1\tilde\lambda_1+\lambda_5\tilde\lambda_5\ ,\quad
P_{\ubar{2}}=\lambda_2\tilde\lambda_2+\lambda_4\tilde\lambda_4\ ,\quad
p_3=\lambda_3\tilde\lambda_3\ .
\end{equation}
Evidently, we have two on-shell mass conditions now:
\begin{equation}
\ket 15 \bra 51 = {m_1}^2 	\ ,		\qquad
\ket 24 \bra 42 = {m_2}^2 	\ . 	\label{mass12}
\end{equation}
The system of LG differential equations for the lowest-spin amplitude $M^{-s_1,-s_2,h_3}\!\equiv\!M$, derived from~\eqref{eqsM3ls}, now contains seven equations. Five of them do not involve the raising operators, and are the following ones:
\begin{equation}
\label{eqs2Mleg}
\begin{aligned}
\left(\lambda_1\frac{\partial}{\partial\lambda_1}-\tilde\lambda_1\frac{\partial}{\partial\tilde\lambda_1}-\lambda_5\frac{\partial}{\partial\lambda_5}+\tilde\lambda_5\frac{\partial}{\partial\tilde\lambda_5}\right) M 
& 	= +2s_1\, M \ , 
&	\left(\lambda_1\frac{\partial}{\partial\lambda_5}-\tilde{\lambda}_5\frac{\partial}{\partial\tilde\lambda_1}\right)M & =0 \ , \\
\left(\lambda_2\frac{\partial}{\partial\lambda_2}-\tilde\lambda_2\frac{\partial}{\partial\tilde{\lambda_2}}-\lambda_4\frac{\partial}{\partial\lambda_4}+\tilde\lambda_4\frac{\partial}{\partial\tilde{\lambda_4}}\right) M 
& 	= +2s_2\, M \ , 
&	\left(\lambda_2\frac{\partial}{\partial\lambda_4}-\tilde{\lambda}_4\frac{\partial}{\partial\tilde\lambda_2}\right)M & =0 \ ,	\\
\left(\lambda_3\frac{\partial}{\partial\lambda_3}-\tilde\lambda_3\frac{\partial}{\partial\tilde\lambda_3}\right)M &= -2h_3\, M \ .
\end{aligned}
\end{equation}
We want again to translate the variables from spinor to spinor products, of which we have now ten angle and ten square ones. Let us sketch how we derive the relations among them. Within the angle products, we choose as independent ones those containing $1$ or $2$. The remaining ones, $\ket34,\ket35$ and $\ket45$, follow from Schouten. Within the square products, using
\begin{equation}
\tilde\lambda_1 = \frac{\ket23}{\ket12}\tilde\lambda_3+\frac{\ket24}{\ket12}\tilde\lambda_4+\frac{\ket25}{\ket12}\tilde\lambda_5 \ ,\quad
\tilde\lambda_2 = -\frac{\ket13}{\ket12}\tilde\lambda_3-\frac{\ket14}{\ket12}\tilde\lambda_4-\frac{\ket15}{\ket12}\tilde\lambda_5 \ ,
\end{equation}
we can express everything in terms of $\bra35$, $\bra43$ and $\bra45$. Using the mass conditions~\eqref{mass12} the following two further relations can be found
\begin{equation}\label{3543}
{m_{1}}^2=-\frac{\ket15\ket23}{\ket12}\bra35-\frac{\ket15\ket24}{\ket12}\bra45 \ ,\quad
{m_{2}}^2=-\frac{\ket31\ket24}{\ket12}\bra34-\frac{\ket15\ket24}{\ket12}\bra45 \ .
\end{equation}
The independent variables we are left with are then seven angle products and one square one, that we name as
\begin{equation}
\begin{array}{llll}
x_1=\ket23 \ , \quad & x_2=\ket31  \ , \quad & x_3=\ket12  \ , \quad & x_4=\ket15  \ , \\
x_5=\ket24 \ , \quad & x_6=\ket25  \ , \quad & x_7=\ket14  \ , \quad & y_8=\bra45 \ .
\end{array}
\end{equation}
Playing around a bit, the system~\eqref{eqs2Mleg} can be simplified to
\begin{equation}
\begin{aligned}
x_1\partial_1 M &= \left(s_{2}-s_{1}-h_{3}\right) M \ , \phantom{\Big|} \quad & 
\partial_6 M &= 0 \ , \\
x_2\partial_2 M &= \left(s_{1}-s_{2}-h_{3}\right) M \ , \phantom{\Big|} \quad & 
\partial_7 M &= 0 \ , \\
\left(x_3\partial_3+y_8\tilde{\partial}_8\right)M & =\left(s_{1}+s_{2}+h_{3}\right)M \ .
\end{aligned}
\end{equation}
Discarding delta-functions, the most general solution of this system is
\begin{equation}
\begin{aligned}
&	
M^{-s_{1},-s_{2},\,h_{3}} = x_1^{s_{2}-s_{1}-h_{3}}\, x_2^{s_{1}-s_{2}-h_{3}}\, x_3^{s_{1}+s_{2}+h_{3}}\: f_2\Big(x_4,\,x_5,\,\frac{y_8}{x_3}\Big) = \\
&	
\;\qquad = {\ket12}^{s_{1}+s_{2}+h_{3}}\: {\ket31}^{s_{1}-s_{2}-h_{3}}\: {\ket23}^{s_{2}-s_{1}-h_{3}}\; f_2\bigg({\ket15},{\ket24},\frac{\bra45}{\ket12}\bigg)	\ ,
\end{aligned}	 	\label{sol2m}
\end{equation}
where $f_2$ is again an undetermined, arbitrary function of its arguments. Again we notice that the first two arguments are dimensionful, so we conveniently rewrite $f_2$ as
\begin{equation}
\label{tf2}
f_2\bigg({\ket15},{\ket24},\frac{\bra45}{\ket12}\bigg)=g\,m_{1}^{1-s_{1}-s_{2}+h_{3}-[g]}
\tilde{f}_2\Bigg(\frac{\ket15}{m_{1}},\frac{\ket24}{m_{2}},\frac{\bra45}{\ket12}\,;\frac{m_{2}}{m_{1}}\Bigg) \ .
\end{equation}
We still have to apply to~\eqref{sol2m} the two remaining LG differential conditions:
\begin{equation}
\label{J++12}
\left(\tilde{J}_+^{\ubar{1}}\right)^{2s_{1}+1}M^{-s_{1},-s_{2},\,h_{3}}=0 \ ,\qquad
\left(\tilde{J}_+^{\ubar{2}}\right)^{2s_{1}+1}M^{-s_{1},-s_{2},\,h_{3}}=0 \ .
\end{equation}
As detailed in the appendix~\ref{raising}, the respective solutions are quite simple:
\begin{align}
	\label{tf2.1}
	\tilde{f}_2&=\sum_{k=0}^{2s_{1}}a_k\left(\frac{\ket15}{m_{1}},\frac{\ket24}{m_{2}}\right)\left(\frac{m_{2}}{m_{1}}+\frac{\ket15}{m_{1}}\frac{\ket24}{m_{2}}\frac{\bra45}{\ket12}\right)^{s_{1}+s_{2}+h_{3}-k} \ , \\
	\label{tf2.2}
	\tilde{f}_2&=\sum_{k=0}^{2s_{2}}b_k\left(\frac{\ket15}{m_{1}},\frac{\ket24}{m_{2}}\right)\left(\frac{m_{1}}{m_{2}}+\frac{\ket15}{m_{1}}\frac{\ket24}{m_{2}}\frac{\bra45}{\ket12}\right)^{s_{1}+s_{2}+h_{3}-k}\ ,
\end{align}
where the $a_k,b_k$ are unknown functions, which we cannot further determine at the moment. We will show how to constrain them later in Section \ref{meet}. We have two expressions for $\tilde{f}_2$ which must be obviously compatible. To analyze this compatibility, we treat separately the cases where the two masses are equal or different. It will be convenient to assume that $s_{1}\leq s_{2}$ for the discussion below.

\paragraph{Different masses.} Comparing the two expressions~\eqref{tf2.1}-\eqref{tf2.2} as power series in the variable $\bra45$ we can easily see that, as it was occurring in~\eqref{h1-h2}, there is a restriction on the helicities and spins of the participating particles:
\begin{equation}
\label{h_3}
h_{3}=-s_{1}-s_{2},-s_{1}-s_{2}+1,\ldots,s_{1}+s_{2}\quad\implies\quad
|h_{3}|\leq s_{1}+s_{2} \ .
\end{equation}
Again, the fact that this process can happen for real kinematics (decay of the more massive particle) allows us to do the same reasoning as for the decay of one massive particle (see below equation~\eqref{h1-h2}), and derive the condition above from a simple physical condition. It is a  non-trivial check of our formalism that these conditions come out automatically from the equations!

The explicit form of the $\tilde{f}_2$ compatible with~\eqref{tf2.1}-\eqref{tf2.2} is easy to obtain, but depends on the exact helicities and spins involved.
Here we just note the number of free functions of $\ket15$ and $\ket24$ which the solution depends on:
\begin{equation}
\#\left(\substack{\textrm{number of free}\\ \textrm{constants}}\right)=\left\{
\begin{array}{lcr}
s_{1}+s_{2}-h_{3}+1	& \textrm{if}	& s_{2}-s_{1}\leq h_{3} \leq s_{1}+s_{2} \\
2s_{1}+1	& \textrm{if}	& s_{1}-s_{2}\leq h_{3} \leq s_{2}-s_{1} \\
s_{1}+s_{2}+h_{3}+1	& \textrm{if}	& -s_{1}-s_{2}\leq h_{3} \leq s_{1}-s_{2} \\
\end{array}\right. \ .
\end{equation}

\paragraph{Equal masses.} In this case it is straightforward to analyze the compatibility condition between~\eqref{tf2.1} and~\eqref{tf2.2}. Recalling that we assumed $s_{1}\leq s_{2}$, and denoting the common mass by $m=m_{1}=m_{2}$, we simply get the following functional form for $\tilde{f}_2$:
\begin{equation}
\label{tf2.m}
\tilde{f}_2=\sum_{k=0}^{2s_{1}}a_k\left(\frac{\ket15}{m},\frac{\ket24}{m}\right)\left(1+\frac{\ket15\ket24}{m^2}\frac{\bra45}{\ket12}\right)^{s_{1}+s_{2}+h_{3}-k} \ ,
\end{equation}
with the $a_k$ unknown functions. Recall that this process cannot happen for real kinematics. Accordingly we do not get constraints on the values of $s_{1}$, $s_{2}$ and $h_{3}$. We notice that this is the same situation as in the massless case, where there was no restriction on the helicities involved in the three-point interaction.


\subsection{Three massive legs}
\label{3m}

For the three-point amplitude involving three massive particles we need six light-like momenta:
\begin{equation}
\label{p123456}
P_{\ubar{1}}=\lambda_1\tilde\lambda_1+\lambda_4\tilde\lambda_4\ ,\quad
P_{\ubar{2}}=\lambda_2\tilde\lambda_2+\lambda_5\tilde\lambda_5\ ,\quad
P_{\ubar{3}}=\lambda_3\tilde\lambda_3+\lambda_6\tilde\lambda_6\ .
\end{equation}
The on-shell mass conditions in spinor formalism are
\begin{equation}
\ket 14 \bra 41 = {m_{1}}^2 	\ ,		\qquad
\ket 25 \bra 52 = {m_{2}}^2 	\ , 	\qquad
\ket 36 \bra 63 = {m_{3}}^2 	\ . 	\label{mass123}
\end{equation} 
The system of LG differential equations for the lowest-spin amplitude $M^{-s_{1},-s_{2},-s_{3}}\!\equiv\!M$, without involving the raising operators, is the following:
\begin{equation}
\label{eqs3Mleg}
\begin{aligned}
\left(\lambda_1\frac{\partial}{\partial\lambda_1}-\tilde\lambda_1\frac{\partial}{\partial\tilde\lambda_1} -\lambda_4\frac{\partial}{\partial\lambda_4}+\tilde\lambda_4\frac{\partial}{\partial\tilde\lambda_4}\right)M &=2s_{1}\, M \ , &
\left(\lambda_1\frac{\partial}{\partial\lambda_4} -\tilde{\lambda}_4\frac{\partial}{\partial\tilde\lambda_1}\right)M &=0 \ , \\
\left(\lambda_2\frac{\partial}{\partial\lambda_2}-\tilde\lambda_2\frac{\partial}{\partial\tilde\lambda_2} -\lambda_5\frac{\partial}{\partial\lambda_5}+\tilde\lambda_5\frac{\partial}{\partial\tilde\lambda_5}\right)M &=2s_{2}\, M \ , &
\left(\lambda_2\frac{\partial}{\partial\lambda_5} -\tilde{\lambda}_5\frac{\partial}{\partial\tilde\lambda_2}\right)M &=0 \ , \\
\left(\lambda_3\frac{\partial}{\partial\lambda_3}-\tilde\lambda_3\frac{\partial}{\partial\tilde\lambda_3} -\lambda_6\frac{\partial}{\partial\lambda_6}+\tilde\lambda_6\frac{\partial}{\partial\tilde\lambda_6}\right)M &=2s_{3}\, M \ , &
\left(\lambda_3\frac{\partial}{\partial\lambda_6}-\tilde{\lambda}_6\frac{\partial}{\partial\tilde\lambda_3}\right)M &=0 \ .
\end{aligned}
\end{equation}
Now we can build thirty spinor products. That is actually a lot, but the procedure described in Section~\ref{kine} ensures that we will be able to reduce them to eleven independent ones, using algebraic and kinematic constraints and the mass conditions~\eqref{mass123}. We choose as independent variables the following ones:
\begin{equation}
\begin{aligned}
x_1&=\ket23 \ , &\quad	x_2&=\ket31 \ , &\quad	x_3&=\ket12 \ , \\
x_4&=\ket14 \ , &\quad	x_5&=\ket25 \ , &\quad 	x_6&=\ket36 \ , \\
x_7&=\ket15 \ , &\quad	x_8&=\ket24 \ , &\quad 	x_9&=\ket26 \ ,
\end{aligned}
\qquad\quad
\begin{aligned}
y_{10}&=\bra45 \ , \\
y_{11}&=\bra46 \ .
\end{aligned}
\end{equation}
The system of LG equations~\eqref{eqs3Mleg}, after all substitutions and some simplifications, is equivalent to
\begin{equation}
\begin{aligned}
x_1\partial_1 M &= \left(s_{2}+s_{3}-s_{1}\right) M \ ,\qquad 	&	  \partial_7 M &= 0 \ ,\\
\left(x_2\partial_2+y_{11}\tilde{\partial}_{11}\right)\! M &= \left(s_{3}+s_{1}-s_{2}\right) M \ ,\qquad	&	\partial_8 M &= 0 \ ,\\
\left(x_3\partial_3+y_{10}\tilde{\partial}_{10}\right)\! M &= \left(s_{1}+s_{2}-s_{3}\right) M \ ,\qquad	& 	\partial_9 M &= 0 \ ,
\end{aligned}
\end{equation}
whose most general solution is
\begin{multline}
	\label{sol3m}
	M^{-s_{1},-s_{2},-s_{3}} = x_1^{s_{2}+s_{3}-s_{1}}\, x_2^{s_{3}+s_{1}-s_{2}}\, x_3^{s_{1}+s_{2}-s_{3}}\: f_3\left(x_4,x_5,x_6,\frac{y_{10}}{x_3},\frac{y_{11}}{x_2}\right) \\
	= {\ket12}^{s_{1}+s_{2}-s_{3}}\: {\ket31}^{s_{3}+s_{1}-s_{2}}\: {\ket23}^{s_{2}+s_{3}-s_{1}}\; f_3\left({\ket14},{\ket25},{\ket36},\frac{{\bra45}}{{\ket12}},\frac{{\bra64}}{{\ket31}}\right) \ ,
\end{multline}
where again $f_3$ is another undetermined function of its arguments. We encounter the same feature of dimensionful arguments as in the two previous cases, so once again we rewrite:
\begin{equation}
\label{tf3}
f_3=g\,m_{1}^{1-s_{1}-s_{2}-s_{3}-[g]}
\tilde{f}_3\left(\frac{\ket14}{m_{1}},\frac{\ket25}{m_{2}},\frac{\ket36}{m_{3}},\frac{{\bra45}}{{\ket12}},\frac{{\bra64}}{{\ket31}}\,;\frac{m_{2}}{m_{1}},\frac{m_{3}}{m_{1}}\right) \ .
\end{equation}
This arbitrary function $f_3$ must be still constrained by the three conditions (one per particle) on the last line of~\eqref{eqsM3ls}. Similarly to the previous subsections, one finds that the solution to the conditions corresponding to particles $2$ and $3$ is
\begin{equation}
\label{tf3.23}
\begin{aligned}
\tilde{f}_3=&\left(1-\frac{m_2}{m_1}\frac{\ket14}{m_1}\frac{\ket25}{m_2}\frac{\bra45}{\ket12}+
\frac{m_3}{m_1}\frac{\ket14}{m_1}\frac{\ket36}{m_3}\frac{\bra64}{\ket31}\right)^{s_1-s_2-s_3}\times\\
&\sum_{n=0}^{2s_2}\sum_{m=0}^{2s_3}a_{n,m}\,\left(\frac{\ket14}{m_1}\frac{\ket25}{m_2}\frac{\bra45}{\ket12}\right)^n\left(\frac{\ket14}{m_1}\frac{\ket36}{m_3}\frac{\bra64}{\ket31}\right)^m\ ,
\end{aligned}
\end{equation}
with $a_{n,m}=a_{n,m}\left(\frac{\ket14}{m_{1}},\frac{\ket25}{m_{2}},\frac{\ket36}{m_{3}}\right)$. This expression can be constrained further with the third requirement:
\begin{equation}
\label{J++11}
\left(\tilde{J}_+^{\ubar{1}}\right)^{2s_{1}+1}M^{-s_{1},-s_{2},-s_{3}}=0 \ .
\end{equation}
Unfortunately the most general solution of this equation is not so simple, depending on the relations that can arise among the three spins and masses. Given the plethora of such cases, and the lesser phenomenological relevance of this type of scattering, we leave a general discussion for future work. Of course any given case can be analyzed with the techniques presented here. For the theoretical purposes that we have in this paper, equation~\eqref{tf3.23} will be enough, as we show in what follows.

\smallskip

\section{Meeting some physical requirements}
\label{meet}

In the previous section we answered the mathematical question of which are the solutions of the system of equations~\eqref{eqsM3ls}. The resulting three-point amplitudes involve the functions $(\tilde{f}_1,\tilde{f}_2,\tilde{f}_3)$, depending on certain combinations of the kinematic variables and which are not fully determined by the equations, similarly to what happened in the massless case (\textit{cf.} equation~\eqref{M3nom}). In that case, imposing some sensible extra physical requirements allowed for a complete specification of the arbitrary function, up to a coupling constant. Let us discuss now similar physical conditions that we can impose on the amplitude for massive scattering, that will allow to strongly constrain $\tilde{f}_1$, $\tilde{f}_2$, and $\tilde{f}_3$.

The first thing that comes to mind is to restrict to real kinematics, $y_i=-{x_i}^*$, as it was done in the massless case. Contrary to that case, this will not get us the job done here, although for certain mass combinations we do obtain some constraints that the arbitrary functions should satisfy. Namely when the three-particle scattering is kinematically prohibited on the real sheet, the amplitude should vanish when $x_i,y_i=0$. This is the case for instance for the scattering of two massive particles of the same mass. This loosely constrains $f_2$. Similar conditions can be derived for $f_3$ when the heaviest mass is smaller than the mass of the other two. However in the cases where the real three-particle scattering is not kinematically prohibited\footnote{These can be interpreted as one massive particle decaying into: two massless particles, a less massive particle plus a massless one, or two less massive particles (their masses summing less than the decaying mass).}, no condition is imposed at all.

\smallskip

A more fruitful condition comes from the following observation. For massive kinematics, we can perform the following transformation:
\begin{equation}
\label{fakeLG}
\quad 	\mu\to t\,\mu \ , \quad 	\tilde\mu\to t^{-1}\tilde\mu \ ,
\qquad	\textrm{when}\quad	P=\lambda\tilde\lambda+\mu\tilde\mu \ .
\qquad	\left(t=e^{i\alpha}\;\;\textrm{\small if }P\in\RR^4\right)
\end{equation}
This transformation changes the kets (bras) involving $\mu$ ($\tilde\mu$), but leaves $P$ invariant. Yet it is not a LG transformation (no $\widetilde{\SU}(2)$ element can generate it via~\eqref{tJs}). It belongs instead to the $\widetilde{\U}(1)\subset\widetilde{\U}(2)$ that we mentioned below~\eqref{cina}. Since this transformation cannot be associated to a LG transformation, it is not physical and one is free to impose whatever transformation property on the amplitude under it (see~\cite{Conde:2016izb} for a more formal justification). In particular, we will require that \textit{the amplitude be invariant} under~\eqref{fakeLG}. This choice is natural for the massless limit that we discuss in the next subsection\footnote{
	In the case we have a Lagrangian, it is also easy to understand this condition. It just follows from the fact that the only dependence that the amplitude can have under~\eqref{fakeLG} appears through polarization tensors. If these are invariant, as it is our case (see Section~\ref{masspol}), the amplitude must be invariant as well.
}.

We can immediately see that invariance of the amplitude under~\eqref{fakeLG} imposes strong constraints on how the functions $\tilde{f}_1,\tilde{f}_2,\tilde{f}_3$ can depend on the ``mass kets''.
Actually, from equations~\eqref{tf1},~\eqref{tf2.1}, \eqref{tf2.2}, \eqref{tf2.m} and~\eqref{tf3.23}, we observe that this invariance requires the arbitrary functions appearing in those equations, depending only on mass kets, to be constant. This fixes completely the functional dependence of $\tilde{f}_1,\tilde{f}_2,\tilde{f}_3$ up to just some constants!

Interestingly, this is essentially the same result that we had for massless particles. But with one important difference: In the massless case there were only two functional structures (\textit{cf.} equations~\eqref{M3H} and~\eqref{M3A}), accompanied by two constants, which can be directly identified with the three-particle coupling constants\footnote{
	Each of these coupling constants corresponds, up to linear combinations, to a parity-even or parity-odd interacion.
}. In the current massive case, the number of functional structures is one in the case of one massive leg, but generically grows with the spin of the massive particles involved if there are more than one of them. The constants accompanying these functional structures will be related to the coupling constants of the theory, but in a theory-dependent way.

\subsection{The massless UV limit}
\label{ssec:UV}

There is one further physical requirement from which we have not profited so far. Generically a massive scattering must look like a massless one in the UV regime\footnote{
	This does not mean that one can always extract massless interactions from massive ones, as this limit might be singular, as it happens for instance for string theory.
}. There is a direct way of implementing this condition in our formalism. Let us rewrite~\eqref{Pmass} as
\begin{equation}
\label{2Elvangframe}
P=p^{\bot}-\frac{m^2}{2P\cdot q}\,q=
\lambda\tilde\lambda+\frac{m^2}{\Pgen{\mu}{P}{\tilde\mu}}\,\mu\tilde\mu \ ,
\end{equation}
where we have to require that the four-vector $q=\mu\tilde\mu$ has a non-vanishing product with $P$. What this choice does is to project massive vectors along a given reference null vector $q$. The massless limit can now be taken simply as $m\to0$ :
\begin{equation}
\label{UVlimit}
P=\lambda\tilde\lambda+\frac{m^2}{\Pgen{\mu}{P}{\tilde\mu}}\,\mu\tilde\mu\quad \xrightarrow[UV limit]{massless} \quad \lambda\tilde\lambda \ .
\end{equation}
After this limit, the massive amplitude should become the massless one. To make this statement concrete, we should identify the map between a massive state $\st{\lambda\tilde\lambda+\mu\tilde\mu\,;\,s,\tilde{\sigma}}$ and a massless one $\st{\lambda\tilde\lambda\,;\,h}$. This is quite simple if we remember the comments we made below equation~\eqref{tJs}, as then it is clear that
\begin{equation}
\st{\lambda\tilde\lambda+\mu\tilde\mu\,;\,s,-s}\quad \xrightarrow[UV limit]{massless} \quad \st{\lambda\tilde\lambda\,;\,-s} \ ,\qquad
\st{\lambda\tilde\lambda+\mu\tilde\mu\,;\,s,s}\quad \xrightarrow[UV limit]{massless} \quad \st{\lambda\tilde\lambda\,;\,s} \ ,
\end{equation}
and the other intermediate spin states decouple. Remembering the identification~\eqref{st=psi}, we can see this reduction explicitly at the level of polarization tensors. For spin $\frac12$, the left-hand side of equation~\eqref{eDirac} trivially reduces to the left-hand side of~\eqref{polsm0}. For spin 1, we see from equation~\eqref{pols1bis} that the neutral polarization vector explodes, and the $\pm$ polarization vectors become the massless ones~\eqref{polsm0}. Interestingly, the reference vector used for projecting the massive momentum becomes the arbitrary vector that appears in~\eqref{polsm0}.

Therefore, if we take the limit~\eqref{UVlimit} in our expressions for the massive lowest-spin amplitudes~\eqref{sol1m},~\eqref{sol2m} and~\eqref{sol3m}, the limiting expression should coincide with~\eqref{M3H} or~\eqref{M3A}. Let us analyze the case in which we have to compare with~\eqref{M3H}\footnote{
	The other case, where the massless amplitude to be matched is~\eqref{M3A}, can be analyzed similarly, only that it is then more convenient to work with with the highest-spin amplitude, which follows from the lowest-spin one just flipping the signs of the spins and interchanging kets by bras.
}.

We write the UV limit condition explicitly for the case of two massive particles, using the expression~\eqref{tf2}, which is valid for both the equal- and different-mass cases. With the identification $h_1=-s_1$, $h_2=-s_2$, the powers involving the angle products match and we are left with
\begin{equation}
\label{match}
g_{\textrm{H}}=g\,{m_1}^{1+h_1+h_2+h_3-[g]}
\tilde{f}_2\left(\frac{\ket{\lambda_1}{\mu_1}}{\sqrt{\Pgen{\mu_1}{P_{\ubar{1}}}{\tilde\mu_1}}},\frac{\ket{\lambda_2}{\mu_2}}{\sqrt{\Pgen{\mu_2}{P_{\ubar{2}}}{\tilde\mu_2}}},0\,;\frac{m_2}{m_1}\right) \ .
\end{equation}
We are now free to perform the transformation~\eqref{fakeLG} in the equation above. The equality can only hold if $\tilde{f}_2$, when the third argument is put to zero, is a constant! Notice that, in particular, this implies that the amplitude must be invariant under~\eqref{fakeLG}, as we assumed in the previous subsection. Moreover, counting mass dimensions in~\eqref{match}, we see that $[g]=1+h_1+h_2+h_3=[g_\textrm{H}]$. The coupling appearing in the lowest-spin massive amplitude, $g$ must be therefore equal to the (holomorphic) coupling of the limiting UV massless, $g_\textrm{H}$, scattering times a dimensionless function of the masses.

Let us further remark that the massless UV limit sets automatically $\bra45=0$. Nevertheless, we can think of imposing this condition in general, outside this limit, and for any process with more than one massive particle. Such a condition has a very natural interpretation: it just means that in the decomposition~\eqref{2Elvangframe} we are choosing the same null reference vector, $q=\mu_q\tilde\mu_q$, for all the massive momenta involved in the scattering process:
\begin{equation}
\label{Elvangframe}
P_{\ubar{i}}=\lambda_i\tilde\lambda_i+\frac{{m_i}^2}{\Pgen{\mu_q}{P_{\ubar{i}}}{\tilde\mu_q}}\,\mu_q\tilde\mu_q \ .
\end{equation}
This way of decomposing massive momenta using a reference vector is actually a very common one~\cite{Cohen:2010mi,Boels:2009bv,Craig:2011ws,Boels:2011zz,Boels:2012if}. If we choose this decomposition, which can be thought simply as a particular frame of our more general construction, the resulting massive amplitudes are quite simple. The line of reasoning leading to~\eqref{match} can be extended straightforwardly to the two other massive cases, leading to the same conclusion: the arbitrary function appearing in~\eqref{sol1m} is constant (in the kinematic variables, it can depend on the masses), the same as the arbitrary functions appearing in equations~\eqref{sol2m} and~\eqref{sol3m} when we go to the frame~\eqref{Elvangframe}. So, in this frame, the lowest-spin massive amplitude takes the universal form
\begin{equation}
\label{theM}
M^{\eta_1,\eta_2,\eta_3}_{\textrm{ls}} = \ket23^{\eta_1-\eta_2-\eta_3}\, \ket31^{\eta_2-\eta_3-\eta_1}\, \ket12^{\eta_3-\eta_1-\eta_2}\,\tilde{g}\big(\{m_i\}\big)\ .
\end{equation}
Although we derived this expression from the massless UV limit, it is immediate that using the frame~\eqref{Elvangframe} in the amplitudes derived in Section~\ref{solving}, together with the invariance under~\eqref{fakeLG}, yields the same result. We stress that, if our physical assumptions hold, this result is non-perturbative, the same as for the massless case~\eqref{M3H}-\eqref{M3A}.


\smallskip

Before moving on to some examples, let us make some remarks here. Notice that in the case we are not in the frame~\eqref{Elvangframe}, while $\tilde{f}_1$ is always a constant (that can be determined from the massless limit), $\tilde{f}_2$ and $\tilde{f}_3$ remain determined only up to some constants. The number of undetermined constants depends on the case under consideration. From imposing the massless limit for both helicities, we can obtain \textit{a priori} $2n$ linear constraints on these constants, if $n$ is the number of massive legs. In certain cases, when the number of free constants is less than $2n$, this might lead to a full determination of the amplitude.

\section{Practical applications}
\label{pract}

In the previous sections we have derived the general form that all three-point scattering amplitudes of any Lorentz-invariant QFT involving massive particles should satisfy. It involves several constants, although under certain circumstances, we can reduce them to a single constant. It is time to test our theoretical predictions, coming from the \mbox{S-matrix} analysis of Lorentz symmetry, with concrete \emph{tree-level} amplitudes computed from the Lagrangian formulation of given QFTs.

Of course we cannot survey all possible QFTs involving massive particles. We will show the match of our formulas with some examples that we consider representatives. As we did in Section~\ref{solving}, we treat the cases with one, two and tree massive particles separately.

\subsection{One massive leg}

This case has been discussed in Section \ref{1m}, and we refer to the notation there. In particular, our conclusion was that the lowest-spin three-point amplitude should look as~\eqref{sol1m}, with the function $f_1$ equal to constant. At tree level, the value of this constant can be determined simply from dimensional analysis:
\begin{equation}
\label{g1m}
f_1=
\begin{cases}
g 	\phantom{\frac{}{|}} 
&	\textrm{if}\quad h_1+h_2-s_3<0 \ , \\
g\,m_3^{1+h_1+h_2-s_3-[g]} 	\phantom{\Big|} 
& 	\textrm{if}\quad h_1+h_2-s_3>0 \ , \\
0 	\phantom{\frac{|}{}} 
& 	\textrm{if}\quad h_1+h_2-s_3=0 \ .
\end{cases}
\end{equation}
In the first line, $g$ is the coupling appearing in the UV-limiting massless amplitude. We have assumed parity invariance in writing $g$ again in the second line. The zero in the third line is due to the non-existence of the massless process with $h_1+h_2-s_3=0$. In the following we show the match with two common processes of LHC physics, plus we comment on how the Landau-Yang theorem follows from our result.

\subsubsection{One massive scalar and two massless vectors}
\label{Hgg}

A prototype of this process is Higgs production via gluon fusion, with a vertex involving two massless gluons; or the decay of a Higgs particle into two photons. Both processes are generated at one loop in the Standard Model, integrating out massive quarks (and also massive vector bosons) for the first (second). For concreteness, let us refer to the first one. Here we consider an effective tree-level vertex. The interaction term in the Standard Model Lagrangian looks like ${\cal L}_{Hgg}\sim H\tr G_{\mu\nu}G^{\mu\nu}$, with coupling constant with mass dimension equal to $-1$.

As a function of the helicity of the gluons, we can in principle consider three configurations. All of them are lowest-spin amplitudes since the Higgs is a scalar. We neglect color labels as they are not relevant for this discussion. The amplitudes can be found for example in~\cite{Dixon:2004za}:
\begin{equation}
M^{-1,-1,0}=g\,\ket12^2 \ , \quad
M^{-1,+1,0}=0 \ , \quad
M^{+1,+1,0}=g\,\bra12^2=g\,m_H^4\frac{1}{\ket12^2} \ .
\end{equation}
The three amplitudes perfectly match our predictions~\eqref{sol1m}, taking~\eqref{g1m} into account. Before moving on to another type of process, we mention that the decay of Higgs to fermion/anti-fermion can be analyzed in a very similar fashion.

\subsubsection{One massive vector and two massless spin-$\frac12$ fermions}

This is a more interesting case than the previous one, given that the massive particle now has spin, so the different possible polarization states come into play. In principle many processes fall into this category, like for example the decays $W^+\to u+\bar{d}$, $Z^0\to q+\bar{q}$, or $W^-\to e^-+\bar{\nu}_e$ at energies where electron and anti-neutrino can be considered massless. Although the Lagrangian vertices for these processes are different, at the level of the three-point amplitude, all those differences just go into the coupling constant (which can carry certain quantum numbers). The kinematic part of the amplitude is the same.

Let us then pick for example the $Z^0\to f+\bar{f}$ process, where $f$ is some generic right-handed Weyl fermion. The Lagrangian vertex reads as ${\cal L}_{Zf\bar{f}}\sim\bar{f}\sigma^{\mu}Z_{\mu} f$, governed by a dimensionless coupling constant. It gives rise to the following three-point amplitudes:
\begin{equation}
\label{Smu}
M^{-\frac12,+\frac12,\tilde{\sigma}_3}=g\,\Pgen{1}{\sigma^{\mu}}{2}\epsilon_{\mu}(P_{\ubar{3}};\tilde{\sigma}_3) \ ,\qquad
M^{+\frac12,-\frac12,\tilde{\sigma}_3}=g\,\Pgen{2}{\sigma^{\mu}}{1}\epsilon_{\mu}(P_{\ubar{3}};\tilde{\sigma}_3) \ ,
\end{equation}
and zero in the configuration where fermion and anti-fermion have the same helicity. In particular, let us focus on the lowest-spin amplitudes. Taking the polarization tensors written in~\eqref{pols1}, and redefining the coupling constant absorbing some numerical factors, we obtain a result that perfectly matches our expectations:
\begin{equation}
M^{-\frac12,+\frac12,-1}=g\,\frac{\ket31^2}{\ket12^{\phantom{2}}} \ ,\quad
M^{\pm\frac12,\pm\frac12,-1}=0 \ ,\quad
M^{+\frac12,-\frac12,-1}=g\,\frac{\ket23^2}{\ket31^{\phantom{2}}} \ .
\label{-1/2+1/2-1}
\end{equation}
For completeness, we display the amplitudes involving the other possible polarization states of the $Z^0$. They can be obtained either directly plugging~\eqref{pols1bis} in~\eqref{Smu}, or by acting with\footnote{
	One should treat with care the existence of the constraints~\eqref{1m.cons}. One can see that $J^{\ubar{3}}_+=-\frac{m}{x_4}\left(\frac{x_3x_4+x_2x_5}{x_1}\partial_2-x_5\partial_1\right)$.
}
$\tilde{J}^{\ubar{3}}_+=\sqrt{-\frac{\bra34}{\ket34}}\left(-\lambda^{(4)}\frac{\partial}{\partial\lambda^{(3)}}+\tilde\lambda^{(3)}\frac{\partial}{\partial\tilde\lambda^{(4)}}\right)$ on~\eqref{-1/2+1/2-1} as instructed by equation~\eqref{eqsM3}. We just display one helicity configuration (swap $1\rightleftharpoons2$ for the other):
\begin{align}
	\label{-1/2,+1/2,0}
	M^{-\frac12,+\frac12,0}=&\frac{1}{\sqrt{2}}\tilde{J}^{\ubar{3}}_+M^{-\frac12,+\frac12,-1}=g\,\sqrt{2}m\,\frac{\ket31\ket14}{\ket12\ket34} \ ,\\
	\label{-1/2,+1/2,+1}
	M^{-\frac12,+\frac12,+1}=&\frac{1}{\sqrt{2}}\tilde{J}^{\ubar{3}}_+M^{-\frac12,+\frac12,0}=g\,m^2\frac{\ket14^2}{\ket12\ket34^2} \ .
\end{align}
Opposite to the lowest-spin amplitude, these ``higher-spin'' amplitudes can -- and do -- depend on $\ket14$. Notice that the dependence is such that invariance under transformation~\eqref{fakeLG} is attained.

\subsubsection{Landau-Yang theorem}

The Landau-Yang theorem~\cite{Landau:1948kw,Yang:1950rg} is an old result about the impossibility of a spin-1 particle to decay into photons. Its phenomenological relevance has been brought up recently by the diphoton events observed at CERN. From our expression for the amplitude with one massive particle~\eqref{sol1m} (with $f_1$ being constant), it is easy to see why such a decay process is prohibited. Putting $|h_1|=|h_2|=s_3=1$, we see that on the one hand, interchanging particles 1 and 2 flips the sign of the amplitude. On the other hand, since the photons are bosons, the amplitude should not be affected by this interchange. The only possible conclusion is that the amplitude be zero. This is just the same reasoning one can use in the massless case to rule out self-interaction among photons (one needs to promote photons to gluons including color labels).

\subsection{Two massive legs}

In this category we find well-known processes like the QED vertex. The notation has been laid out in Section \ref{2m}, and here we will match~\eqref{tf2.m} as we will only treat cases with two equal-mass particles, $m_1=m_2=m$. Moreover we will check that the three-point amplitude becomes of the form~\eqref{theM} when $\bra45=0$, with the function $\tilde{g}$ in~\eqref{theM} fixed, by dimensional analysis at tree level, to
\begin{equation}
\label{g2m}
\tilde{g}(m)=
\begin{cases}
g & \textrm{if}\quad s_1+s_2-h_3>0 \ , \\
g\,m^{1-s_1-s_2+h_3-[g]} & \textrm{if}\quad s_1+s_2-h_3<0 \ , \\
0 & \textrm{if}\quad s_1+s_2-h_3=0 \ .
\end{cases}
\end{equation}
Let us illustrate our predictions with three examples.

\subsubsection{Two massive scalars and a massless vector}

Such an amplitude occurs naturally for example when computing rational pieces of one-loop YM amplitudes. If one uses supersymmetry as a technical trick, then a scalar loop must be evaluated in $4+2\varepsilon$ dimensions, so that effectively one can think of computing a four-dimensional amplitude where the scalars are massive.

The amplitudes with two massive scalars plus a massless gluon can be readily found in helicity-spinor form in~\cite{Badger:2005zh}. Using the kinematic constraints to express the amplitudes in terms of the spinor variables that we have considered independent, we can write them as
\begin{equation}\begin{aligned}
&M^{0,0,-}=g\,\frac{\ket23\ket31}{\ket12}\left(1+\frac{\ket15\ket24}{m^2}\frac{\bra45}{\ket12}\right)^{-1} \ ,\\
&M^{0,0,+}=g\,m^2\,\frac{\ket12}{\ket23\ket31}\left(1+\frac{\ket15\ket24}{m^2}\frac{\bra45}{\ket12}\right) \ .
\end{aligned}\end{equation}
They match perfectly the expectation~\eqref{tf2.m}, where a single constant enters in the amplitude, which must therefore coincide with the coupling constant up to powers of the masses. These powers are as in~\eqref{g2m}, and by putting $\bra45=0$ we see that we recover~\eqref{theM}, as we predicted.

\subsubsection{Two massive and one massless vectors}

In the Coulomb branch of $\nn=4$ Super Yang-Mills we can find a coupling between a gluon and two vector bosons. Such type of interactions have been cunningly studied by Craig et al. in~\cite{Craig:2011ws} (see also~\cite{Kiermaier:2011cr}), where they found that the lowest-spin three-point amplitude involving a positive-helicity gluon looks like:
\begin{equation}
M^{-1,-1,+1}=g\,\frac{\ket12^3}{\ket23\ket31}\left(1+\frac{\ket15\ket24}{m^2}\frac{\bra45}{\ket12}\right) \ .
\label{-1,-1,+1}
\end{equation}
This certainly matches~\eqref{tf2.m} with $a_0=0=a_1$, and $a_2=1$. Moreover, in the $\bra45=0$ frame~\eqref{theM} is recovered with the function $\tilde{g}$ given by~\eqref{g2m}. A further check of our results that we can do is to confront the ``higher-spin'' amplitudes, $M^{0,-1,+1}$ and $M^{+1,-1,+1}$ that we can obtain by acting with $J^{\ubar{1}}_+$ on~\eqref{-1,-1,+1}, with the ones written in~\cite{Craig:2011ws}, computed with different methods. We obtain a perfect match. Here we are more interested in pointing out a curious case, that of the all-minus amplitude $M^{-1,-1,-1}$.

Our expectation is that $M^{-1,-1,-1}=0$ in the $\bra45=0$ frame. Should this imply that it be zero in a different Lorentz frame? The answer is no! This amplitude has been computed in~\cite{Craig:2011ws}, and it casts as
\begin{equation}
\label{-1,-1,-1}
M^{-1,-1,-1}=\frac{g}{m^2}\,\ket12\ket23\ket31\,\frac{\left(\frac{\ket15}{m}\right)^2\left(\frac{\ket24}{m}\right)^2\left(\frac{\bra45}{\ket12}\right)^2}{1+\frac{\ket15}{m}\frac{\ket24}{m}\frac{\bra45}{\ket12}} \ .
\end{equation}
Interestingly, this functional form can be seen to come from~\eqref{tf2.m} by appropriately tuning the three constants that appear in this case, namely $a_0=1=a_2$, $a_1=-2$. Of course, the combination is such that it collapses to zero when $\bra45=0$.

\subsubsection{Two massive fermions and one massless vector. QED}

In the QED that we know and love we can find two different vertices coupling a photon with an electron/positron pair. One is the electric coupling and the other the magnetic dipole coupling, ${\cal L}_{e^-\!e^+\!\gamma}=\bar{\psi}\left(e\,\gamma^{\mu} A_{\mu}+\frac{g}{2}\,\gamma^{\mu\nu}F_{\mu\nu}\right)\psi$. With this parametrization, $e$ is a dimensionless coupling constant while $g$ has negative mass dimension, $[g]=-1$. We can immediately compute the three-point
lowest-spin amplitude as
\begin{equation}
\label{-1/2,-1/2,-1}
\begin{aligned}
M^{-\frac12,-\frac12,-1}&=\bar{v}_-(P_{\ubar{1}})\left(e\,\gamma^{\mu}+\frac{g}{2}\,\gamma^{\mu\nu}p^{3}_{\nu}\right)\epsilon_{\mu}(p_3;-1)\,u_-(P_{\ubar{2}}) \\
&=\ket23\ket31\left(\frac{e}{m}\,\frac{\frac{\ket15\ket24}{m^2}\frac{\bra45}{\ket12}}{1+\frac{\ket15\ket24}{m^2}\frac{\bra45}{\ket12}}+g\right)
\end{aligned}
\end{equation}
where we have used the polarization vector from~\eqref{polsm0}, and the Dirac spinors
\begin{equation}
\bar{v}_-(P_{\ubar{1}})=\left(\,\sqrt{\frac{\ket15}{\bra51}}\,\tilde\lambda_5\; \lambda_1\,\right) \ ,\qquad
u_-(P_{\ubar{2}})=\left(\!\!\begin{array}{c}-\sqrt{\frac{\ket24}{\bra42}}\,\tilde\lambda_4 \\ \lambda_2 \end{array}\!\!\right) \ .
\end{equation}
Looking at equation~\eqref{-1/2,-1/2,-1}, we get again a match with our predictions. For the electric coupling, we have to choose $a_0=1=-a_1$ in~\eqref{tf2.m}, while for the magnetic coupling we should put $a_0=1$, $a_1=0$. So in this case, we see that the two functional structures appearing in the three-point amplitude are possible.

We can also check the predictions in the frame $\bra45=0$ frame. Given that the for the electric coupling the helicity configuration $(-\frac12,-\frac12,-1)$ is not allowed in the massless UV limit, we expect a null scattering amplitude in the special frame; and this is indeed the case. Instead for the magnetic coupling we can reach such a configuration. Therefore we expect the functional form~\eqref{theM}, with the function $\tilde{g}$ equal to the coupling constant; and this is precisely what happens in~\eqref{-1/2,-1/2,-1}.

\subsection{Three massive legs}

This case was analyzed in Section~\ref{3m}, where we arrived at the solution~\eqref{sol3m}, with the function $\tilde{f}_3$ given by~\eqref{tf3.23}. This solution can be constrained further with condition~\eqref{J++11}, whose consequences we chose not to work out in general. We will show how that can be done with one particular example, where two masses are the same: the electroweak decay of a Z-boson: $Z^{0}\to b+\bar{b}$. Other processes with the same spin structure as this one would be QED with a Proca Lagrangian, or another electroweak decay: $W^-\to e^-+\bar{\nu}_e$. The latter would in principle involve three different masses, but since neutrino masses are subtle, we prefer to discuss the Z-boson decay.

Let us first compute the three-point amplitude of the process with the well-known methods. The Lagrangian three-point vertex for the electroweak decay of the $Z^0$ looks like ${\cal L}_{Zb\bar{b}}\sim\,Z_{\mu}\,\bar{b}\left(g_V\gamma^{\mu}+g_A\gamma^{\mu}\gamma^5\right)b$, with $[g_V]=[g_A]=0$. Identifying the Z-boson, $\bar{b}$ antiquark and $b$ quark as particles $1,2,3$ respectively, we can use the following polarization tensors for the lowest-spin states:
\begin{equation}
\epsilon_-(P_{\ubar{1}})=\frac{\lambda_1\tilde\lambda_4}{\bra14} \ ,\qquad
\bar{v}_-(P_{\ubar{2}})=\left(\,\sqrt{\frac{\ket25}{\bra52}}\,\tilde\lambda_5\; \lambda_2\,\right) \ ,\qquad
u_-(P_{\ubar{3}})=\left(\!\!\begin{array}{c}-\sqrt{\frac{\ket36}{\bra62}}\,\tilde\lambda_6 \\ \lambda_3 \end{array}\!\!\right) \ .
\end{equation}
Neglecting parity violation, \textit{i.e.} setting $g_A=-g_V=g_{W}/2$, the computation of the three-point amplitude easily yields
\begin{equation}
\label{-1,-1/2,-1/2.3m}
\begin{aligned}
M^{-1,-\frac12,-\frac12}&=\frac{g_W}{2}\,\bar{v}_-(P_{\ubar{2}})\gamma^{\mu}\left(1-\gamma_5\right)\epsilon_{\mu}(P_{\ubar{1}};-1)\,u_+(P_{\ubar{3}}) \\
&=\frac{g_W}{m_2}\ket12\ket31\frac{\bra45}{\ket12}\frac{\ket14}{m_1}\frac{\ket25}{m_2} \ .
\end{aligned}
\end{equation}
We now want see if this expression matches the one we can derive with our methods. For this particular configuration\footnote{
	Had we labeled the vector boson as particle 3 and the two quarks as particles 1 and 2, condition~\eqref{J++11} would no longer be trivial, and it would eliminate two of the six constants appearing in~\eqref{tf3.23}. Of course the final result for the lowest-spin three-point amplitude does not depend on how we label the particles.
}, it turns out that~\eqref{J++11} is automatically satisfied by the function~\eqref{tf3.23} put in~\eqref{sol3m}. So, the generic form of the three-point amplitude that we can derive is
\begin{multline}
	\label{-1,-1/2,-1/2.3m.th}
	M_3^{-1,-\frac12,-\frac12}=g\,{m_1}^{-1-[g]}\ket12\ket31\left(a_{0,0}+a_{1,0}\frac{\ket14}{m_1}\frac{\ket25}{m_2}\frac{\bra45}{\ket12}\right.\\
	\left.+a_{0,1}\frac{\ket14}{m_1}\frac{\ket36}{m_3}\frac{\bra64}{\ket31}+
	a_{1,1}\frac{\ket14^2}{m_1^2}\frac{\ket25}{m_2}\frac{\ket36}{m_3}\frac{\bra45}{\ket12}\frac{\bra64}{\ket31}\right) \ .
\end{multline}
In order to match~\eqref{-1,-1/2,-1/2.3m}, we see that it is enough to choose as the only non-zero coefficient $a_{1,0}=\frac{m_1}{m_2}$. Finally let us notice that in the special frame $\bra45=0=\bra64$ the amplitude vanishes, as we could have predicted because otherwise it should be equal to $g_\textrm{W}$, which does not have the proper dimensions.

\section{Looking backward, and forward}
\label{back}

Let us start our conclusions by summarizing the results of this text. In any QFT containing only massless particles and enjoying Poincar\'e symmetry, the three-point scattering amplitude can be determined from very general assumptions, namely Poincar\'e symmetry itself and a few sensible physical requirements. Motivated by this fact, we have investigated whether this property holds in the case where some of the particles are massive. 

The helicity-spinor formalism is a convenient (not essential) tool for analyzing the LG scaling equations associated to Lorentz symmetry. That was already well established in the massless case, but it turns out to be true also when massive momenta are involved. We implemented the formalism representing a massive momentum in terms of two massless ones as in equation~\eqref{Pmass}. Then we were able to derive the system of differential equations~\eqref{eqsM3} that the massive amplitude should satisfy. These equations are slightly more involved than their massless counterparts, but they can still be solved, determining part of the functional form of the massive three-point amplitude, but leaving certain arbitrariness.

By imposing some physical requirements on the amplitude, we were able to reduce the arbitrary functions just to constants, exactly as it happens for the massless three-point amplitude. Namely, one of the conditions is the invariance of the amplitude under the transformation~\eqref{fakeLG}, that we can think as complementary to the LG equations. The freedom to impose this condition essentially corresponds to a choice of polarization tensors, and in Section \ref{ssec:UV} we showed that such a choice is the natural one from the perspective of the massless limit (for large momenta, the amplitude involving massive particles should reduce to the massless one). Contrary to the massless case, the constants that appear in the massive three-point amplitude cannot always be fixed beforehand from the coupling constants of the three-particle interaction, only in some low-spin cases.

In practice, one typically chooses a common reference vector to define asymptotic massive states, \textit{i.e.} fixing $\mu\tilde\mu$ in equation~\eqref{Elvangframe}. The massive three-point amplitudes simplify further when this choice is made. In particular, the lowest-spin amplitude turns out to be pretty similar to the massless one, and can be fixed up to a coupling constant. We finally contrasted in Section \ref{pract} our results with some amplitudes that can be found in the literature, computed with usual Lagrangian methods.

\medskip

We believe that the current results could serve as a starting point for several ideas that come to mind. First it seems worth to investigate further the connection explored in Section \ref{ssec:UV}, where we observed the following phenomenon occurring when taking the massless UV limit: the spinors making up the reference null vector for a massive momentum, as in~\eqref{Elvangframe}, become the arbitrary spinors that appear in the polarization vector of the limiting massless spin-$1$ particle. One may then foresee the intriguing possibility of accommodating the Higgs mechanism within this formalism, generalising the ideas in~\cite{Craig:2011ws,Kiermaier:2011cr}.

Another straightforward direction is to use the massive three-point amplitude as a building block for constructing higher-point amplitudes, for instance by means of the BCFW recursion relations. A promising avenue is that of the four-particle test~\cite{Benincasa:2007xk,Benincasa:2011pg} (see also~\cite{McGady:2013sga}), which consists in constructing four-particle amplitudes out of three-particle ones, using different BCFW shifts and requiring independence of the used shift. This imposes strong constraints on the possible interactions already for massless particles, and it would be interesting to see if the addition of mass brings something new to the table. It seems reasonable to believe that the four-particle test could help in determining the free constants that appear in the massive amplitudes. It would be nice to see that interactions among massive high-spin particles are much less constrained than their massless counterparts. A particularly fascinating case of study would be that of gravitons plus massive higher-spin states.


Finally, it could also be interesting to understand if our results can have a higher-di\-men\-sional origin. Via Kaluza-Klein reduction of some higher-dimensional theory containing only massless particles, it is possible to generate interactions among massless and massive particles in four dimensions (see~\cite{Bern:2010qa,Elvang:2011fx,Davies:2011vt,Craig:2011ws} for some concrete examples). This would lead us to investigate the uniqueness of three-point massless interactions in higher-dimensional theories (in six dimensions some partial results have been already obtained in~\cite{Cheung:2009dc,Czech:2011dk}).

\section*{Acknowledgments}

We are especially grateful to Riccardo Argurio for stimulating comments throughout the development of this work, and on the final manuscript. We have also benefited from conversations with Robson Alves Do Nascimento Filho, Laure-Anne Douxchamps, Yu-tin Huang, Gustavo Lucena G\'omez, Euihun Joung, Karapet Mkrtchyan, Micha Moskovic, Diego Redigolo and Dario Rosa in the early and late stages of this work. Useful comments from Paolo Benincasa, Euihun Joung, and Diego Redigolo have improved the quality of the text. It is a pleasure to thank them all. The work of EC was supported in part by the National Research Foundation of Korea through the grant NRF-2014R1A6A3A04056670, and the grants 2005-0093843, 2010-220-C00003 and 2012K2A1A9055280. The work of AM is supported by IISN-Belgium (convention 4.4503.15).

\appendix

\section{Details on the action of the Little Group}
\label{LG}

This appendix is devoted to an explicit study of the action of the Lorentz group in spinor language. It contains no new results; its purpose is to re-derive known formulas in our notation for the convenience of some readers.

For starters, we work out the exact form of $\zeta(\Lambda)$ in~\eqref{Ltransf} when $\Lambda$ is a spatial rotation around one of the coordinate axis. Let us for example consider a rotation of angle $\theta$ around $x^3$. The generator of this transformation on four-vectors is $\hat{J}^{12}$:
\begin{equation}
R_3(\theta)=e^{i\theta\hat{J}^{12}}=\left(
\begin{array}{cccc} 1&0&0&0\\0&\cos(\theta)&-\sin(\theta)&0\\0&\sin(\theta)&\cos(\theta)&0\\0&0&0&1
\end{array}\right)\;.
\end{equation}
On the matrix $k_{a\dot{a}}$ the action will be:
\begin{equation}
\left(\begin{array}{cc} k_0+k_3 & k_1-ik_2\\k_1+ik_2 & k_0-k_3\end{array}\right)\mapsto\left(\begin{array}{cc} k_0+k_3 & e^{-i\theta}(k_1-ik_2)\\e^{i\theta}(k_1+ik_2) & k_0-k_3\end{array}\right) \ ,
\end{equation}
and the $SL(2,\mathbb{C})$ elements which implement this transformation are:
\begin{equation}
\zeta\left(e^{i\theta\hat{J}^{12}}\right)=
\pm\left(\begin{array}{cc} e^{-i\theta/2} & 0\\0 & e^{i\theta/2}\end{array}\right)=\pm e^{-i\frac{\theta}{2}\sigma^3} \ .
\label{rot3}
\end{equation}
The rotations around the other two axis are as one could have expected: $\zeta\left(e^{i\theta\hat{J}^{23}}\right)=\pm e^{-i\frac{\theta}{2}\sigma^1}$, $\zeta\left(e^{i\theta\hat{J}^{31}}\right)=\pm e^{-i\frac{\theta}{2}\sigma^2}$. A rotation $R$ in a generic direction can be obtained in several ways\footnote{Like for instance via Euler angles: $R(\phi,\theta,\psi)=R_3(\psi)R_1(\theta)R_3(\phi)$ $\implies$ $\zeta(R)=e^{-i\frac{\psi}{2}\sigma^3}e^{-i\frac{\theta}{2}\sigma^1}e^{-i\frac{\phi}{2}\sigma^3}$.}, and the result is always an $\SU(2)$ matrix
\begin{equation}\label{SU2g}
\zeta(R)=\pm\left(\begin{array}{cc} a & b \\ -b^{*} & a^{*} \end{array}\right)\;,\qquad |a|^2+|b|^2=1 \ .
\end{equation}
In the following let us just keep one element in the equivalence class  $\SU(2)/\ZZ_2$, and suppress the $\pm$ signs. The eigenvectors of the matrix give the direction of rotation, and the eigenvalues $e^{\pm i\theta/2}$ the angle of rotation $\theta$. One can check  this fact with an explicit parametrization of $R$.

If spatial rotations build up an $\SU(2)$ subgroup inside $\SL(2,\mathbb{C})$, boosts must be in equivalence with the quotient $\SL(2,\mathbb{C})/\SU(2)$. One can work out the spinor form of a boost exactly as done for the rotation. Take for instance a boost along the $x^3$ direction, which has the following four-vector form:
\begin{equation}
B_3(\beta)=e^{i\beta \hat{J}^{03}}=\left(
\begin{array}{cccc} \cosh(\beta)&0&0&-\sinh(\beta)\\0&0&0&0\\0&0&0&0\\-\sinh(\beta)&0&0&\cosh(\beta)
\end{array}\right)\;.
\end{equation}
The only components changing are $p_0+p_3\to e^{-\beta}(p_0+p_3)$, $p_0-p_3\to e^{\beta}(p_0-p_3)$, so the associated two-by-two matrix is
\begin{equation}\label{boost3}
\zeta(e^{i\beta \hat{J}^{03}})=
\left(\begin{array}{cc} e^{-\beta/2} & 0\\0 & e^{\beta/2}\end{array}\right)=e^{-\frac{\beta}{2}\sigma^3} \ .
\end{equation}
This is a suggestive form, and one can check that for a boost of rapidity $\beta$ along the $x^i$ direction, the associated matrix turns out to be $\zeta(e^{i\beta \hat{J}^{0i}})=e^{-\frac{\beta}{2}\sigma^i}$. However, given that we know how a generic rotation looks like, the form of~\eqref{boost3} is enough to write down a generic boost. It suffices to use the matrix $R$ that takes the $x^3$ direction into the direction of interest, then the boost of rapidity $\beta$ along this direction is just obtained as
\begin{equation}
B(\beta)=R\,B_3(\beta)\,R^{-1}\quad\implies\quad \zeta(B(\beta))=\zeta(R)\zeta(B_3(\beta))\zeta(R^{-1}) \ .
\end{equation}

\subsection{Massless Little Group}
\label{LG.1}

Let us take a null momentum $p=\lambda\tilde\lambda$. The group of Lorentz rotations that leave it invariant is isomorphic to $\ISO(2)$. If we do not want to involve boosts, this group is reduced to $\U(1)$. This restriction is the one we adopted in Section \ref{BC}. The $\U(1)$ LG is clearly made of the rotations around the spatial direction $\vec{p}$. Let us find out how it acts on the spinors $\lambda$, $\tilde\lambda$.

It is enough to remember the comments below $\eqref{SU2g}$. For a rotation $R$ of angle $\theta$ around the $\vec{p}$ direction, the corresponding matrix acting on the spinors is $\zeta(R)$. Since $\lambda$ is precisely an eigenvector of $\zeta(R)$ with eigenvalue $\theta/2$, the action of the rotation on the spinors is straightforward.
\begin{equation}\label{Ronl}
\zeta(R)\lambda=e^{-i\theta/2}\,\lambda \ ,\qquad\qquad\tilde\lambda\,\zeta(R)^\dagger=e^{i\theta/2}\tilde\lambda \ ,
\end{equation}
where we are assuming that $p$ is real, $\tilde\lambda=\sign(p_0)\lambda^\dagger$. For the complex case the LG would be~$\mathbb{C}$, with the transformation $\lambda\to t\lambda$, $\tilde\lambda\to t^{-1}\tilde\lambda$ with $t\in\mathbb{C}$, which obviously reduces to~\eqref{Ronl} applying the real condition.

For the sake of concreteness, let us illustrate equation~\eqref{Ronl} with one particular example. First let us write explicitly
\begin{equation}\label{mus}
\lambda=\left(\!\begin{array}{c}\lambda_1\\ \lambda_2\end{array}\!\right) \ ,\quad\tilde{\lambda}=\left(\,\tilde{\lambda}_{\dot{1}}\;\,\tilde{\lambda}_{\dot{2}}\,\right) \ ,\qquad
\lambda\tilde{\lambda}=
\left(\begin{array}{cc} \lambda_1\tilde{\lambda}_{\dot{1}} & \lambda_1\tilde{\lambda}_{\dot{2}} \\ \lambda_2\tilde{\lambda}_{\dot{1}} & \lambda_2\tilde{\lambda}_{\dot{2}}\end{array}\right)=
\left(\begin{array}{cc} p_0+p_3 & p_1-ip_2 \\ p_1-ip_2 & p_0-p_3
\end{array}\right) \ .
\end{equation}
Take a massless particle moving along the $x^3$ direction, with momentum $p_{\mu}=(\frac12,0,0,\frac12)$ in some units. We can choose the following spinorial representation of such a momentum
\begin{equation}
\lambda=\left(\!\begin{array}{c}1\\ 0\end{array}\!\right) \ ,\quad\tilde{\lambda}=\Big(\,1\,\;0\,\Big) \ ,\qquad\qquad
\lambda\tilde{\lambda}=\left(\begin{array}{cc} 1 & 0 \\ 0 & 0 \end{array}\right) \ .
\end{equation}
The generator of the $\U(1)$ LG must be identified with $R=e^{i\theta \hat{J}^{12}}$. From equation~\eqref{rot3}, we know that the corresponding matrix acting on the spinors is $e^{-i\frac{\theta}{2}\sigma^3}$. Although such a matrix will not act diagonally on a generic spinor, its action on these particular $\lambda$, $\tilde\lambda$ is indeed diagonal:
\begin{equation}
e^{-i\frac{\theta}{2}\sigma^3}\,\lambda=e^{-i\theta/2}\lambda \ ,\quad\qquad \lambda^{\dagger}e^{i\frac{\theta}{2}\sigma^3}=e^{i\theta/2}\tilde\lambda \ ,
\label{LGaction}
\end{equation}
as we wanted to see.

\subsection{Massive Little Group}
\label{LG.2}

Let us now consider a massive momentum, that we write as $P=\lambda\tilde\lambda+\mu\tilde\mu$. In the rest frame, where $P_{\mu}=(m,0,0,0)$, things are particularly simple, because the LG is directly identified with the $\SO(3)$ of spatial rotations, and we learnt already how these act on spinors. In a generic frame we just need to conjugate with the boost that takes us to the rest frame. That is, if a given frame is a boosted version of the rest frame via a boost $B$, the LG elements in such a frame will be simply
\begin{equation}
B\,R\,B^{-1} \ ,
\end{equation}
with $R$ a spatial rotation. The action of this LG element on the spinors that make up $P$ is as follows:
\begin{equation}\label{Btransf}
\begin{aligned}
&&\lambda\to \zeta(B)\zeta(R)\zeta(B^{-1})\,\lambda \ ,\quad&&\tilde\lambda\to \tilde\lambda\,\zeta^{\dagger}(B^{-1})\zeta^{\dagger}(R)\zeta^{\dagger}(B) \ ,\\
&&\quad\mu\to \zeta(B)\zeta(R)\zeta(B^{-1})\,\mu \ ,\quad&&\tilde\mu\to \tilde\mu\,\zeta^{\dagger}(B^{-1})\zeta^{\dagger}(R)\zeta^{\dagger}(B) \ .
\end{aligned}
\end{equation}
Although it is perfectly fine, this is not a very beautiful action. The main inconvenience for us is that the spinors do not scale homogeneously as in the massless case~\eqref{Ronl}, but instead their two components scale independently.

To illustrate this point more clearly, let us work out a particular example. The inconvenience happens already in the case where we are in the rest frame, $B=1$. If we take a rotation around the $x^3$ axis, generated by $J_0=\hat{J}^{12}$, the matrix to be used in~\eqref{Btransf} has been already written in~\eqref{rot3}. Looking at the infinitesimal version of~\eqref{Btransf} for this rotation, we can immediately write the expression of $J_0$ in terms of spinor differentials as
\begin{equation}
J_0=-\frac12\left(\lambda_1\frac{\partial}{\partial\lambda_1}-\lambda_2\frac{\partial}{\partial\lambda_2}-\tilde\lambda_1\frac{\partial}{\partial\tilde\lambda_1}+\tilde\lambda_2\frac{\partial}{\partial\tilde\lambda_2}+\mu_1\frac{\partial}{\partial\mu_1}-\mu_2\frac{\partial}{\partial\mu_2}-\tilde\mu_1\frac{\partial}{\partial\tilde\mu_1}+\tilde\mu_2\frac{\partial}{\partial\tilde\mu_2}\right) \ .
\end{equation}
Clearly such an operator would not act nicely on the spinor products like $\ket{\lambda}{\kappa}$, $\bra{\tilde\mu}{\tilde\kappa}$, etc.

The inconvenience can be resolved if we recast transformation~\eqref{Btransf} as~\eqref{cina}, that we rewrite here.
\begin{equation}\label{cina.app}
\left(\!\begin{array}{c}\lambda\\ \mu\end{array}\!\right)\to U\left(\!\begin{array}{c}\lambda\\ \mu\end{array}\!\right) \ ,\qquad
\left(\,\tilde\lambda \;\; \tilde\mu\,\right) \to \left(\,\tilde\lambda \;\; \tilde\mu\,\right)U^{\dagger} \ ,
\end{equation}
Clearly, after quotienting by $\ZZ_2$, the degrees of freedom in this transformation are the same as those of the $\SO(3)$ rotations. But, is it possible to find a map between~\eqref{Btransf} and~\eqref{cina.app}? The answer is yes.

To simplify the upcoming expressions, we omit the $\zeta(\cdot)$ notation of~\eqref{Btransf}, and assume that $R$ and $B$ stand already for the two-by-two matrices acting on the spinors. The first step to find the map is to show that, for real momenta, the equation
\begin{equation}\label{eqBU}
\left(\begin{array}{c}BRB^{-1}\,\lambda\\ BRB^{-1}\,\mu\end{array}\right)= U\left(\begin{array}{c}\lambda\\ \mu\end{array}\right)\ ,
\end{equation}
implies the ``conjugate''
\begin{equation}\label{eqBUt}
\left(\;\tilde\lambda(BRB^{-1})^{\dagger}\;\;\tilde\mu(BRB^{-1})^{\dagger}\;\right)=\left(\,\tilde\lambda \;\; \tilde\mu\;\right)U^{\dagger} \ .
\end{equation}
To do that, we need the conditions that follow from expanding in components
\begin{equation}\label{BmB}
(\lambda\tilde\lambda+\mu\tilde\mu)=B\left(\begin{array}{cc} m & 0 \\ 0 & m \end{array}\right)B^\dagger \ ,
\end{equation}
with the specific form of the boost $B$. Complex-conjugating the component equations of~\eqref{eqBU}, and using~\eqref{BmB} plus the reality conditions $\tilde\lambda=\pm\lambda^\dagger$ and $\tilde\mu=\pm\mu^\dagger$, we directly get the component equations of~\eqref{eqBUt}. Then we arrive to the conclusion that it is enough to solve~\eqref{eqBU}. If we parametrize the matrices $R$ and $U$ as
\begin{equation}
R=\left(\begin{array}{cc} a & b \\ -b^{*} & a^{*} \end{array}\right)\;,\quad |a|^2+|b|^2=1\quad;\quad
U=\left(\begin{array}{cc} \alpha & \beta \\ -\beta^{*} & \alpha^{*} \end{array}\right)\;,\quad |\alpha|^2+|\beta|^2=1 \ ,
\end{equation}
what we want is a map $(a,b)\mapsto(\alpha,\beta)$. Such a map is easily found to be
\begin{equation}
\label{map}
\begin{array}{cc}
\displaystyle	\alpha = \frac{\ket{BRB^{-1}\lambda}{\mu}}{\ket{\lambda}{\mu}} \ ,		& \displaystyle	\beta = \frac{\ket{\lambda}{BRB^{-1}\lambda}}{\ket{\lambda}{\mu}} \ ,	\\
\displaystyle	\alpha^{*} = \frac{\ket{\lambda}{BRB^{-1}\mu}}{\ket{\lambda}{\mu}} \ ,	& \displaystyle	\beta^{*} = \frac{\ket{\mu}{BRB^{-1}\mu}}{\ket{\lambda}{\mu}} \ .
\end{array}
\end{equation}
Notice that the condition $|\alpha|^{2}\!+\!|\beta|^{2}\!=\!1$ will be preserved in virtue of $|\det(R)|\!=\!1$. This is the map that translates an $\SU(2)$ transformation into an $\widetilde{\SU}(2)$ one.

\section{Details on the action of raising operators}
\label{raising}

In Section~\ref{solving} we have worked out how to solve the system of LG differential equations in order to get the generic form of a three-point amplitude under the assumption of Poincar\'e invariance. We have then applied an additional constraint, coming from the action of the spin-raising operator, namely the last equation of the system~\eqref{eqsM3ls}. However, for the sake of readability, we have skipped there the details of such action, which we report in this appendix. We follow the division of the three cases exactly as in Section~\ref{solving}. Several of the algebraic manipulations performed in this appendix assume implicitly momentum conservation and Schouten identities.

\subsection{One-massive amplitude}

We first rewrite the expression~\eqref{sol1m} in the following way:
\begin{align}
	M^{h_{1},\,h_{2},-s_{3}} &
	=	\left(\frac{\ket23}{\ket31\ket12}\right)^{h_1} \left(\frac{\ket31}{\ket12\ket23}\right)^{h_2} \left(\frac{\ket12\ket34}{\ket23\ket31}\right)^{-s_3} {\ket34}^{s_3}\, f_1\big(\ket34\big) \nonumber \\
	&
	=	{\Ac_1}^{h_1}\, {\Ac_2}^{h_2}\, {\Ac_3}^{-s_3}\; F_1(\mu_3)\, \phantom{\frac{|}{}}	,
\end{align}
defining
\begin{eqnarray}
&\displaystyle	\Ac_1 = \frac{\ket23}{\ket31\ket12} \ ,		\qquad
\Ac_2 = \frac{\ket31}{\ket12\ket23} \ , 		\qquad
\Ac_3 = \frac{\ket12\ket34}{\ket23\ket31} \ ;		& \\
& 	F_1(\mu_3) \equiv {\mu_3}^{s_3}\: f_1(\mu_3)  \ , \quad \text{with }\ \mu_3\equiv\ket34 \ . & \phantom{\frac{|}{|}}
\end{eqnarray}
The operator that raises the spin component of the massive particle is defined in the first line of equation~\eqref{tJs}, and in the current case it reads\footnote{
	We have decided to omit the scaling factors of~\eqref{rJ+} in this appendix as the resulting expressions are simpler in this way. Since these factors are completely transparent to the action of the raising operators, it is trivial to reinstate them back.
}
\begin{equation}
J_+^{\ubar{3}}= -\lambda_4\partial_3 +\tilde{\lambda}_3\tilde{\partial}_4 \ .
\label{J+3}
\end{equation}
It acts only on the~$\Ac_i$, as follows:
\begin{equation}
J_+^{\ubar{3}} \Ac_1 = +\Ac_1\Ac_3 \ , \quad 	J_+^{\ubar{3}} \Ac_2 = -\Ac_2\Ac_3 \ , \quad 	J_+^{\ubar{3}} \Ac_3 = \Ac_3\left(\Ac_3+2b\right) \ ;
\end{equation}
with
\begin{equation}
b=\frac{\ket24}{\ket23} \ ,	\qquad 	J_+^{\ubar{3}}b=b^2 \ .
\end{equation} 
Then the action of $J_+^{\ubar{3}}$ on the amplitude $M=M^{h_{1},\,h_{2},-s_{3}}$ yields
\begin{equation}
J_+^{\ubar{3}}M = M\left( \Delta\Ac_3 -2s_3b \right) \ , \qquad \text{with }\, \Delta=h_1-h_2-s_3 \ ,
\end{equation}
and acting repeatedly we obtain
\begin{equation}
(J_+^{\ubar{3}})^n M = M\; \sum_{k=0}^{n} {n\choose k} (\Delta)_{(k)} (-2s_3+k)_{(n-k)}\; {\Ac_3}^k \, b^{n-k} \ , \label{J3nM1}
\end{equation}
where ${n}\choose{k}$ are the usual binomial coefficients, and $(a)_{(n)}$ is the raising Pochhammer symbol~\eqref{Pochhammer}. The formula~\eqref{J3nM1} can be proved by induction.
%
Then it is easy to extract from the form the formula~\eqref{J3nM1} the desired constraint on the helicities. Indeed, when $n\!=2s_3\!+\!1$ (\emph{i.e.} when $J_+^{\ubar 3}$ annihilates the amplitude), since $(-a)_{(a+1)}\!=\!0$ except for $a\!+\!1\!=\!0$, only the term for $k\!=\!n$ survives, yielding the condition
\begin{equation}
(\Delta)_{(2s_3+1)}=0 \ ,
\end{equation}
which is eventually the condition~\eqref{J++1m}.

\subsection{Two-massive amplitude}

We analogously rewrite the expression~\eqref{sol2m} as:
\begin{align}
	M^{-s_1,-s_2,\,h_3} &
	=	\left(\frac{\ket23}{\ket31\ket12}\right)^{-s_1} \left(\frac{\ket31}{\ket12\ket23}\right)^{-s_2} \left(\frac{\ket12}{\ket23\ket31}\right)^{h_3}\, f_2\Big(\ket14,\ket25;\,\frac{\bra54}{\ket12}\Big) \nonumber \\
	&
	=	{\Ac_1}^{-s_1}\, {\Ac_2}^{-s_2}\, {\Ac_3}^{h_3}\; F_2\big(\mu_1,\mu_2;\,\xi\big)\, \phantom{\frac{|}{}}	,
\end{align}
defining
\begin{eqnarray}
&\displaystyle	\Ac_1 = \frac{\ket23\ket15}{\ket31\ket12} \ ,		\qquad
\Ac_2 = \frac{\ket31\ket24}{\ket12\ket23} \ , 		\qquad
\Ac_3 = \frac{\ket12}{\ket23\ket31} \ ;		& \\
& 	F_2(\mu_1,\mu_2;\,\xi) \equiv {\mu_1}^{s_1}{\mu_2}^{s_2} f_2(\mu_1,\mu_2;\,\xi)  \ , \quad \text{with }\ \mu_1\equiv\ket15\ ,\, \ \mu_2\equiv\ket24\ ,\, \ \xi=\frac{\bra54}{\ket12} \ . & \phantom{\frac{|}{|}}
\end{eqnarray}
In this case we shall consider the action of the spin-raising operator for both massive particles, $i\!=1,2$, which reads as:
\begin{equation}
\begin{aligned}
J_+^{\ubar{i}} \Ac_i & = 				\big(\epsilon_i\Ac_i+2b_i\big)\, \Ac_i 	\ , \\
J_+^{\ubar{i}} \Ac_{\slashed i} & = 		+\epsilon_i\Ac_i\,\Ac_{\slashed i} 	\ , \\
J_+^{\ubar{i}} \Ac_1 & =					-\epsilon_i\Ac_i\,\Ac_3 	\ ,
\end{aligned}
\end{equation}
where $\epsilon_i\!\equiv {\slashed i}\!-\!i$ is just a sign. In addition, we have now the dependence of the function~$F_2$ on the variable~$\xi$, which is not invariant under the action of the spin-raising operators:
\begin{equation*}
	J_+^{\ubar{i}}\xi= \epsilon_i\Ac_i\: x_i \ , \quad \text{with }\ x_i= \xi -\frac{{m_{\slashed i}}^2}{\mu_1\mu_2}\ .
\end{equation*}
Despite of that, the algebraic structure is very similar to the one-massive case, and the final expression can again be proved by induction, even if now it involves derivatives of the function~$F_2$. It reads
\begin{align}
	&\!\!\!	{\Ac_1}^{s_1}{\Ac_2}^{s_2}{\Ac_3}^{-h_3}\left(J_+^{\ubar{i}}\right)^n M^{-s_1,-s_2,\,h_3} = \nonumber \\
	& \qquad\qquad =	
	\sum_{k=0}^{n}{n\choose k} x_i^k F_2^{(k)} \sum_{j=k}^{n}{{n-k}\choose{j-k}} (\Delta+k)_{(j-k)} (-2s_i+j)_{(n-j)} \left(\epsilon_i\Ac_i\right)^j b_i^{n-j} \ ,
\end{align}
where $F_2^{(k)}\equiv(\partial_\xi)^kF_2(\xi)$, and $\Delta=-s_1-s_2-h_3\,$.

We notice that also in this case, when we take $n=2s_i+1$ the only term that survives in the second sum is the one for $j\!=\!n$, yielding a differential equation of the \mbox{$2s_i+1$-th} order for $F_2$:
\begin{equation}
\label{JinM2}
(J_+^{\ubar{i}})^{2s_i+\!1} M^{-s_1,-s_2,\,h_3}=0 \quad\Leftrightarrow\quad
\sum_{k=0}^{n}{{n}\choose{k}} (\Delta+k)_{(n-k)}\; x_i^k F_2^{(k)} =0 \ .
\end{equation}
The most general solution of such equation is
\begin{equation}
F_2\big(\ket15,\ket24;\,\xi\big)= x_i^{s_1+s_2+h_3}\; \sum_{k=0}^{2s_i} c^{(i)}_k(\ket15,\ket24)\; x_i^{-k} \ , 	\label{F2xi}
\end{equation}
which, up to redefinition of the coefficients in the sum, gives the expressions~\eqref{tf2.1} and~\eqref{tf2.2}.

\subsection{Three-massive amplitude}

The expression~\eqref{sol3m} for the amplitude involving three massive particles can be recast as well in this useful way:
\begin{equation}
M^{-s_1,-s_2,-s_3}= {\Ac_1}^{-s_1} {\Ac_2}^{-s_2} {\Ac_3}^{-s_3}\: F_3\big(\mu_1,\mu_2,\mu_3;\,\xi_2,\xi_3\big) \ ,
\end{equation}
through the redefinitions
\begin{eqnarray}
&\displaystyle	\Ac_1 = \frac{\ket23\ket14}{\ket31\ket12} \ ,		\qquad
\Ac_2 = \frac{\ket31\ket25}{\ket12\ket23} \ , 		\qquad
\Ac_3 = \frac{\ket12\ket36}{\ket23\ket31} \ ;		& \\
&\displaystyle	\xi_2=\frac{\bra45}{\ket12}	\ , 	\qquad	\xi_3=\frac{\bra64}{\ket31} \ ;	\\
& 	F_3(\cdots;\xi_2,\xi_3) = \mu_1^{s_1}\mu_2^{s_2}\mu_3^{s_3}\: f_3(\cdots;\xi_2,\xi_3) \ , \quad \mu_i\equiv\ket{i}{i\!+\!3} \ . & \phantom{\frac{:}{}}
\end{eqnarray}
We notice that in this case, with the choice of independent variables we have made to derive the expression~\eqref{sol3m}, we have that $J_+^{\ubar{2}}$ and $J_+^{\ubar{3}}$ act only on the respective $\xi_i$, whereas $J_+^{\ubar{1}}$ acts on both of them. So it is easier to consider first the action of $J_+^{\ubar{2}}$ and $J_+^{\ubar{3}}$, which is completely analogous to the one of the previous section, and it yields, for $i=2,3$,
\begin{equation}
J_+^{\ubar{i}}\xi_i=-\epsilon_i\Ac_i\, x_i \ , \qquad 
\text{with }\ x_i\equiv\frac{x}{\mu_i}\ , \quad 
x =\mu_2\xi_2+\mu_3\xi_3 -\frac{{m_1}^2}{\mu_1}\ , 	\label{Jixii}
\end{equation}
and $\epsilon_i\!={\slashed i}\!-\!i$ just a sign.
From this we can derive, again by induction,
\begin{align}
	&\!\!\!\!\!	{\Ac_1}^{s_1} {\Ac_2}^{s_2} {\Ac_3}^{s_3}\, (J_+^{\ubar{i}})^nM^{-s_1,-s_2,-s_3} = 	\nonumber \\
	& \qquad =
	\sum_{k=0}^{n}{n\choose k} (-x_i)^k F_3^{(i,k)}\, \sum_{j=k}^{n}{{n-k}\choose{j-k}} (\Delta+k)_{(j-k)} (-2s_i+j)_{(n-j)}\, (\epsilon_i\Ac_i)^j \, b_i^{n-j} \ , \label{JinM3}
\end{align}
where we have denoted by $F_3^{(i,k)}$ the \mbox{$k$-th} derivative of $F_3$ with respect to $\xi_i$, and
\begin{equation*}
	\Delta=s_1-s_2-s_3 \ ;	\qquad	b_2=\frac{\ket15}{\ket12}\ ,	\quad	b_3=\frac{\ket61}{\ket31}\ .
\end{equation*}
For $n=2s_i\!+\!1$, only the term for $j=n$ survives in the second sum, and this yields basically the same equation as~\eqref{JinM2} for $F_3$, whose most general solution is
\begin{equation}
F_3\big(\mu_1,\mu_2,\mu_3;\,\xi_2,\xi_3\big)=
x^\Delta\: \sum_{k=0}^{2s_i} c^{(i)}_k\big(\mu_1,\mu_2,\mu_3;\xi_{\slashed i}\big)\; x^k \ , 	\label{F3x}
\end{equation}
where we have absorbed the factors~$\mu_2,\mu_3$ into the coefficients~$c_k^{(i)}$, in order to express the result in the unique variable~$x$.

The action of $J_+^{\ubar{1}}$ is quite involved, since it acts on both variables $\xi_2$ and~$\xi_3$, and so in particular also on the coefficients~$c_k^{(i)}(\xi_{\slashed i})$. But working out some simple cases it seems quite likely that this third condition, \emph{i.e.} $(J_+^{\ubar{1}})^{2s_1\!+\!1}M^{-s_1,-s_2,-s_3}=0$, would yield a condition on the spins similar to those in~\eqref{h1-h2} or~\eqref{h_3}, following from conservation of angular momentum. 

%

\bibliography{amplitudesrefs,localrefs}
\bibliographystyle{utphys}
 
\end{document}